\documentclass[12pt,preprint]{aastex}

\usepackage{natbib}

\slugcomment{Submitted to the Astrophysical Journal}

\shorttitle{Multiwavelength properties of X-ray sources}
\shortauthors{Treister et al.}

\begin{document}

\title{The Cal\'an-Yale Deep Extragalactic Research (CYDER) Survey: Optical Properties and Deep Spectroscopy of Serendipitous X-ray Sources\footnote{Partly based on observations collected at the European 
Southern Observatory, Chile, under programs 68.A-0459 and
72.A-0509}}

\author{Ezequiel Treister\altaffilmark{1,2,3}, Francisco J. Castander\altaffilmark{4}, Thomas J. Maccarone\altaffilmark{5}, Eric Gawiser\altaffilmark{1,3,6}, Paolo S. Coppi\altaffilmark{1},  C. Megan Urry\altaffilmark{2}, Jos\'e Maza\altaffilmark{3}, David Herrera\altaffilmark{1}, Valentino Gonzalez\altaffilmark{3}, Carlos Montoya\altaffilmark{3} and  Pedro Pineda\altaffilmark{3}}

\altaffiltext{1}{Department of Astronomy, Yale University, P.O. Box 208101,New Haven,CT 06520.}
\altaffiltext{2}{Yale Center for Astronomy \& Astrophysics, Yale University, 
P.O. Box 208121, New Haven, CT 06520}
\altaffiltext{3}{Departamento de Astronom\'{\i}a, Universidad de Chile, Casilla 36-D, Santiago, Chile.}
\altaffiltext{4}{Institut d'Estudis Espacials de Catalunya/CSIC,Gran Capit\`a 2-4, E-08034 Barcelona, Spain.}
\altaffiltext{5}{Astronomical Institute ``Anton Pannekoek'',University of Amsterdam,1098 SJ Amsterdam, The Netherlands}
\altaffiltext{6}{NSF Astronomy and Astrophysics Postdoctoral Fellow}

\email{treister@astro.yale.edu}

\begin{abstract}
We present the first results from the Cal\'an-Yale Deep
Extragalactic Research (CYDER) survey. The main goal of this
survey is to study serendipitous X-ray sources detected by
$Chandra$ in an intermediate flux range
($10^{-15}-10^{-12}$~ergs~s$^{-1}$) that comprises most of
the X-ray background. 267 X-ray sources spread over 5
archived fields were detected. The $\log N-\log S$
distribution obtained for this sample is consistent with the
results of other surveys. Deep $V$ and $I$ images were taken
of these fields in order to calculate X-ray-to-optical flux
ratios. Identifications and redshifts were obtained for 106
sources using optical spectroscopy from 8-m class telescopes
to reach the optically faintest sources, to the same level
as deeper X-ray fields like the Chandra Deep Fields, showing
that the nature of sources detected depends mostly on the
optical limit for spectroscopy. In general, sources
optically classified as obscured Active Galactic Nuclei
(AGNs) have redder optical colors than unobscured AGN.  A
rough correlation between $f_X/f_{\rm opt}$ and hard X-ray
luminosity was found for obscured AGN confirming the
prediction by existing models that in obscured AGN the
optical light is completely dominated by the host
galaxy. The previously claimed decrease of the obscured to
unobscured AGN ratio with increasing X-ray luminosity is
observed. However, this correlation can be explained as a
selection effect caused by the lower optical flux of
obscured AGN. Comparison between the observed $N_H$
distribution and predictions by existing models shows that
the sample appears complete up to $N_H<3\times
10^{22}$~cm$^{-2}$, while for more obscured sources
incompleteness plays an important role in the observed
obscured to unobscured AGN ratio.
\end{abstract}


\section{Introduction}
\label{intro}

Wide-area X-ray surveys have played a key role in understanding the
nature of the sources that populate the X-ray universe.  Early surveys
like the $Einstein$ Medium Sensitivity Survey \citep{gioia90}, ROSAT
(Roentgen Satellite) International X-ray/Optical Survey
\citep{ciliegi97} and the ASCA (Advanced Satellite for Cosmology and
Astrophysics) Large Sky Survey \citep{akiyama00} showed that the vast
majority of the X-ray sources were AGN. In particular, in shallow wide
area surveys in the soft (0.5-2 keV) X-ray band, most of the sources
detected are unobscured, broad line AGN, which are characterized by a
soft X-ray spectrum with a photon index\footnote{We define the photon
index $\Gamma$ as the exponent giving a photon flux density X-ray
spectrum $dN/dE\propto E^{-\Gamma}$ in photons
cm$^{-2}$~s$^{-1}$~keV$^{-1}$} $\Gamma=1.9$ \citep{nandra94}.

More recent, deeper observations, mostly by ROSAT
\citep{hasinger98}, XMM-Newton and $Chandra$ \citep{rosati02}, that
resolved between 70\% and 90\% of the X-ray background (XRB)
showed that the vast majority of this background radiation
can be attributed to AGN. However, the spectrum of the XRB
is well characterized up to $E\sim30$~keV by a power law
with photon index $\Gamma=1.4$ \citep{gruber99}, harder than
the typical unobscured AGN spectrum
\citep{mushotzky00}. Given that photoelectric extinction
preferentially absorbs soft X-ray photons
\citep{morrison83}, the X-ray spectra of obscured AGN look
harder and therefore more compatible with the observed
spectral shape of the XRB. Therefore, population synthesis
models \citep{madau94,comastri95,gilli99,gilli01} that can
explain the spectral shape and normalization of the XRB use
a combination of obscured and unobscured AGN as the major
contributor. In these models, the ratio of obscured to
unobscured AGN is about 4:1 \citep{gilli01} with a redshift
peak at $z\sim 1.3$. However, recent deep optical
spectroscopic follow up in the $Chandra$ Deep Fields (CDF)
North \citep{barger03} and South \citep{szokoly04} revealed
a much lower redshift peak at $z\sim 0.8$ and an obscured to
unobscured AGN ratio of $\sim$ 2:1.

While large observational efforts have been concentrated in
the Chandra Deep Fields, which provide the deepest view of
the X-ray Universe (e.g., a flux limit of $\simeq 2.5\times
10^{-17}$~ergs~cm$^{-2}$s$^{-1}$ on the CDF-N), the small
area covered ($\simeq 0.07$ deg$^{2}$ each) does not allow
them to obtain a statistically significant number of sources
in the intermediate X-ray flux range ($10^{-15}-10^{-12}$
ergs~cm$^{-2}$~s$^{-1}$) that contributes $\sim$ 60-70\% of
the XRB. Therefore, we obtained identifications and studied
the multiwavelength properties of X-ray sources in this flux
range over a much larger area.

Specifically, in 2001 we started the Cal\'an-Yale Deep Extragalactic
Research (CYDER) survey, a multiwavelength study of serendipitous
X-ray sources in existing, archived, moderately deep $Chandra$
fields. Initial results from the first two fields studied were
presented by \citet{castander03b}. Also, two high-redshift ($z>4$)
X-ray selected quasars discovered in this survey, a significant
fraction of the total sample known today ($\sim$10), were reported by
\citet{castander03a} and \citet{treister04a}. Near infrared images in
the $J$ and $K$ bands were obtained for these fields up to $J\sim 21$
and $K\sim 20$ mags (Vega). The results of combining X-ray/optical and
near infrared observations for our sample of serendipitous X-ray
sources will be presented in a following paper (F. Castander et al, in
prep).

In this paper, we present optical photometry for 267 X-ray
sources selected in the $Chandra$ total band (0.5-8 keV) in
the five fields studied by the CYDER survey. Also,
spectroscopic identifications and redshifts for 106 X-ray
sources are presented. The sample presented here is
comparable in multiwavelength follow-up to deeper, more
famous, surveys like the CDFs and the Lockman
Hole. Spectroscopic identifications were obtained for
sources with relatively faint optical fluxes ($V\sim 24$
mag), allowing for a more unbiased study of the X-ray
population and showing that the statistical properties of
the sample depends significantly on the depth of the
spectroscopic follow-up. Also, the use of five different
fields spread over the sky allows to reduce the effects of
cosmic variance, that affected the results of single-field
studies, e.g., the presence of clusters in the CDF-S
\citep{gilli03}.

In \S~\ref{xray_data} we explain the criteria used to select
the X-ray fields studied and the procedures followed to
reduce the X-ray data and to extract sources lists. In
\S~\ref{opt_data} we describe the optical imaging and
spectroscopy observations and the data reduction methods
used. In \S~\ref{source_prop} we present source properties
in each wavelength range. Our results are discussed in
\S~\ref{disc} and the conclusions outlined in
\S~\ref{conc}. Throughout this paper we assume
$H_o=70$~kms$^{-1}$Mpc$^{-1}$, $\Omega_m=0.3$ and
$\Omega_\Lambda=0.7$, consistent with the cosmological
parameters reported by \citet{spergel03}.

\section{X-ray Data}
\label{xray_data}

\subsection{Field Selection}

Fields in the CYDER survey were selected based on existing
deep $Chandra$ observations available in the public archive
before the optical imaging campaign started in late
2000. The fields are observable from the southern
hemisphere, are at high galactic latitude ($|b|>40^\circ$)
in order to minimize dust extinction, avoiding targeting
known clusters given the difficulties in dealing with a
diffuse non-uniform background. Only observations with the
Advanced CCD Imaging Spectrometer (ACIS;
\citealp{garmire03}) were used. In the case of ACIS-I
observations all four CCDs were used while for observations
in the ACIS-S mode only the S3 and S4 chips were used in
order to keep the off-axis angle small and therefore only
use zones with good sensitivity. Fields selected for this
study are presented in Table \ref{fields}. In two of these
fields, C2 and D2, the original target of the observation
was a galaxy group. In these cases, the galaxy group diffuse
emission reduces the sensitivity in the centers of the
regions, but not dramatically. Roughly the central $40''$
radius region for HCG 62 has substantial gas emission, while
the $1'$ central region was affected by the presence of HCG
90. This accounts for about 2\% of the HCG 62 region and a
slightly smaller fraction of the HCG 90 region, since the
latter was observed with ACIS-I. The effective area of field
C5 was set to 0 since the Chandra images of that field were
read in subraster mode to include only the central source,
and therefore serendipitous sources detected on that field
were ignored to compute the $\log N-\log S$ relation. Some
of our fields were also studied by other similar surveys. In
particular, the Q2345 and SBS 0335 fields were analyzed by
the Chandra Multiwavelength Project (ChaMP; \citealp{kim04})
while the HCG 62 and Q2345 fields were studied by the
Serendipitous Extragalactic X-ray Source Identification
Program (SEXSI; \citealp{harrison03}).

\subsection{X-ray Data Reduction}

Reduction of the data included the removal of bad columns
and pixels using the guidelines specified on the ``ACIS
Recipes: Clean the Data'' web page and the removal of
flaring pixels using the FLAGFLARE routine. We used the full
set of standard event grades (0,2,3,4,6) and created two
images, one from 0.5 to 2.0 keV and one from 2.0 to 8.0
keV. Then, we used the WAVDETECT routine from the CIAO
package to identify the point sources within these images,
checking wavelet scales 1,2,4,8 and 16. Sources were
extracted independently in the soft (0.5-2.0 keV) and hard
(2.0-8.0 keV) band images. The false source detection
probability is set to $10^{-6}$ for ACIS-S observations and
$10^{-7}$ for ACIS-I observations.  This gives a likelihood
of $\sim 1$ false source detection per field observed. Given
the low density of X-ray sources and the good spatial
resolution of Chandra, matching sources in the soft and hard
bands was straightforward.

Where the X-ray spectrum had at least 60 counts, the photons were
binned in groups of 20, and the spectrum was fit in XSPEC 11.0, using
a model consisting of a power law with the appropriate Galactic
absorption value for each field.  Where the number of counts was
smaller than 60, the same procedure was used, except that the spectral
index was fixed to $\Gamma=$1.7, consistent with the hardening of the
X-ray spectrum with decreasing flux \citep{giacconi01}.

\section{Optical Data}
\label{opt_data}

\subsection{Optical Imaging}

Optical images were obtained using the CTIO 4-m Blanco
Telescope in Chile with the MOSAIC-II camera
\citep{muller98}, which has a field of view of $36'\times
36'$. Details of the optical observations are presented in Table
\ref{opt_sum}. All the fields have optical coverage in the $V$ and $I$
filters in the optical, and were also imaged in the $J$ and $K$ bands
in the near infrared, observations that will be reported in a
following publication.

Reduction of the data was performed using standard
procedures included in IRAF v 2.12\footnote{IRAF is
distributed by the National Optical Astronomy Observatory,
which is operated by the Association of Universities for
Research in Astronomy, Inc., under cooperative agreement
with the National Science Foundation.}, in particular in the
MSCRED package. The data reduction scheme followed was based
on the recipe used by the NOAO Deep Wide Survey\footnote{The
data reduction cookbook and a description of the survey
design can be found in the web page
http://www.noao.edu/noao/noaodeep/ReductionOpt/frames.html}.
Standard calibration images (bias and dome flats) were
obtained each night for every filter used. Super-sky flats
were constructed based on several ($\sim$ 20) unregistered
frames in each filter, masking real objects in order to
obtain a secondary flat field image. Once basic calibrations
were performed, individual frames were registered and
co-added to obtain the final image in each
filter. Astrometric solutions for each final image were
calculated based on the USNO catalog. Typical astrometric
uncertainties are $\sim 0.3$'', smaller than the on-axis PSF
size of $Chandra$ images, therefore allowing for an accurate
match between optical and X-ray data.

Objects in the final images were extracted using SExtractor
\citep{bertin96}. To detect objects we used a threshold of
1.5$\sigma$ above the background per pixel and a minimum
area of 15 pixels ($\sim 1.0$~arcsec$^2$) above that
threshold. In several experiments, this combination of
parameters gave a good balance between completeness and
false detections, the latter being lower than $\sim$ 5\%,
for the range of FWHM of our images.

Zero points for each image were obtained independently for
each night based on observations of Landolt standard fields
\citep{landolt92}. Aperture photometry was then performed
using a diameter of 1.4 times the FWHM. Magnitudes were later
corrected for the (small) effects of Galactic extinction in our
high-galactic latitude fields. Limiting magnitude for each image was
calculated based on global RMS measurements of the background and
reported in Table
\ref{opt_sum}.

Objects in the $V$ and $I$ images were matched by position allowing a
maximum distance of 1$''$ between objects in different filters. The
typical difference between the $V$ and $I$ counterparts is $\sim
0.5''$, consistent with the previously reported astrometric
uncertainties and typical centroid errors, so the choice of 1$''$ as a
threshold provides a good balance in order to avoid spurious
matches. If more than one match was found inside that area then the
closest match was used, however this only happened in a few cases
given the typical sky density of our optical images. $V-I$ color was
calculated for sources detected in both bands.

\subsection{Optical Spectroscopy}

Given that one of the goals of CYDER is the study of the
optically faint X-ray population, only 8-m class telescopes
were used for the spectroscopic follow-up. Multi-object
spectrographs were used in order to improve the efficiency
of the observations, including FORS2 at the VLT in MXU mode
and the LDSS-2 instrument on the Magellan I (Baade)
Telescope. Details of these observations are given in Table
\ref{spec_sum}.

Given the space density of X-ray sources at our flux limit
and the field of view of the instruments used, typically
$\sim$8 X-ray sources were observed per mask. For the
observations with FORS2 at the VLT, the 300V-20 grism was
used, which gives a resolution $R\sim 520$ (10.5\AA) for our
1'' slits, with a typical wavelength coverage from
$4000-9000$\AA~depending on the position of the source in
the mask. Observations with LDSS-2 used the Med/Blue grism,
giving a dispersion of 5.3\AA~pixel$^{-1}$ at a central
wavelength of 5500\AA~and resolution of $R\sim$350 with our
1.0'' wide slits. The typical wavelength coverage with this
configuration was $\sim 4000-7500$\AA~.

Spectral reduction was performed using standard IRAF tasks
called by a customized version of the BOGUS
code\footnote{Available at
http://zwolfkinder.jpl.nasa.gov/$\sim$stern/homepage/bogus.html}.
We calibrated the wavelength of the spectrum using He-Ar
comparison lamps and the night-sky lines. In order to
flux-calibrate our spectra, $\sim 2$ spectrophotometric
standards were observed every night.

\subsection{Catalog}

The full catalog of X-ray sources in the CYDER field is presented in
the on-line version of the journal, while for clarity a fraction of
the catalog is presented in Table~\ref{catalog}. The full catalog is
also available on-line at http://www.astro.yale.edu/treister/cyder/.
Coordinates are given as measured in the X-ray image, while the offset
is calculated with respect to the closest optical counterpart and only
reported when this offset is smaller than 2.5'' and thus that
counterpart was used in the analysis. When a counterpart was not
detected in the optical images, the $5\sigma$ upper limit in that band
is reported. In order to convert count rates into fluxes, the
procedure described in section \S2.2 was followed. The observed X-ray
luminosity was computed only for sources with spectroscopic
identification and measured redshift. This luminosity was calculated
in the observed frame without accounting for k-corrections or
correcting for absorption. Therefore, the simple formula $L_X=4\pi
d^2f_X$ was used.

\section{Source Properties}
\label{source_prop}

\subsection{X-ray}

A total of 37 X-ray sources were detected in the C2 field,
with a total (0.5-8 keV) X-ray flux ranging from $1.7\times
10^{-15}$ to $5\times 10^{-13}$~ergs~cm$^{-2}$~s$^{-1}$. In
the C5 field only 5 X-ray sources were detected above
$1.6\times 10^{-14}$~ergs~cm$^{-2}$~s$^{-1}$ (0.5-8 keV)
since only the s3 CCD was read and in subraster mode. In the
D1, D2 and D3 fields 93, 47 and 85 X-ray sources were
detected respectively, with a total (0.5-8 keV) X-ray flux
ranging from $8\times 10^{-16}$ to $1\times
10^{-13}$~ergs~cm$^{-2}$~s$^{-1}$.

\begin{figure}
\figurenum{1}
\includegraphics[angle=0,scale=0.7]{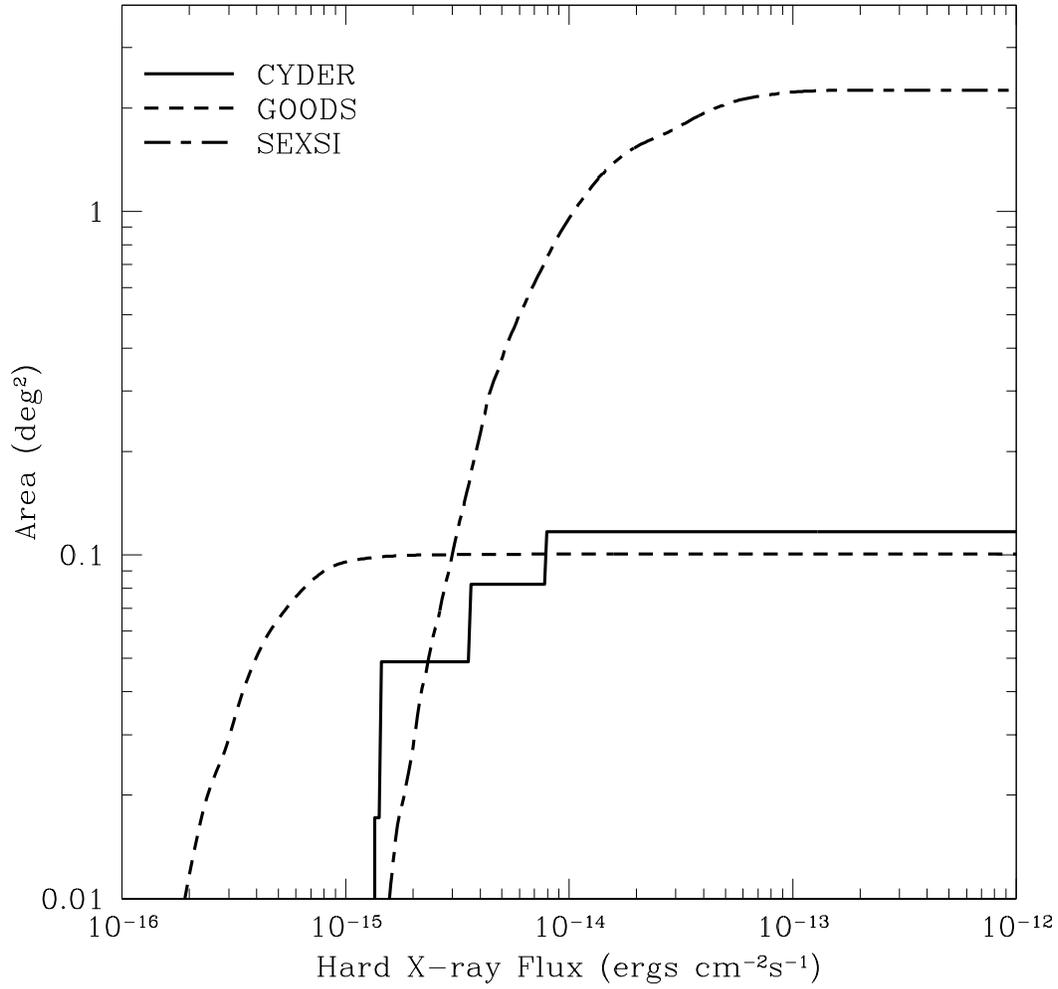}
\caption{Area covered as a function of limiting flux in the hard 
X-ray band for the CYDER survey, compared to other X-ray surveys like
SEXSI \citep{harrison03} and GOODS \citep{alexander03}.}
\label{area}
\end{figure}

The area covered as a function of limiting flux in the hard
X-ray flux band was first estimated individually for each
field using the Portable, Interactive Multi-Mission
Simulator (PIMMS; \citealp{mukai93}). Given the complexities
associated with the modeling of the varying PSF as a
function of off-axis angle and the presence of diffuse
emission in most of these fields that makes the problem of
estimating the completeness levels even harder a constant,
higher, flux limit was used for each field. Specifically,
for each field we used a fixed value of 2.5 times the flux
of the faintest source included in the catalog for each
field, in order to be sure that the sample is complete up to
that flux. This roughly corresponds to 20 counts detected in
the hard band for ACIS-S observations and 10 counts for
fields observed with the ACIS-I CCDs. The flux limit in the
hard band assumed for each field is shown on Table
\ref{fields}. The resulting total area of the survey is $\ga
0.1$deg$^{-2}$, with a minimum flux limit of $1.3\times
10^{-15}$~ergs~cm$^{-2}$~s$^{-1}$ (2-8
keV). Figure~\ref{area} shows the resulting area versus flux
limit curve, in comparison to other surveys like the Great
Observatories Origin Deep Survey (GOODS;
\citealp{giavalisco04}) and SEXSI \citep{harrison03}.

\begin{figure}[!ht]
\figurenum{2}
\includegraphics[angle=0,scale=0.7]{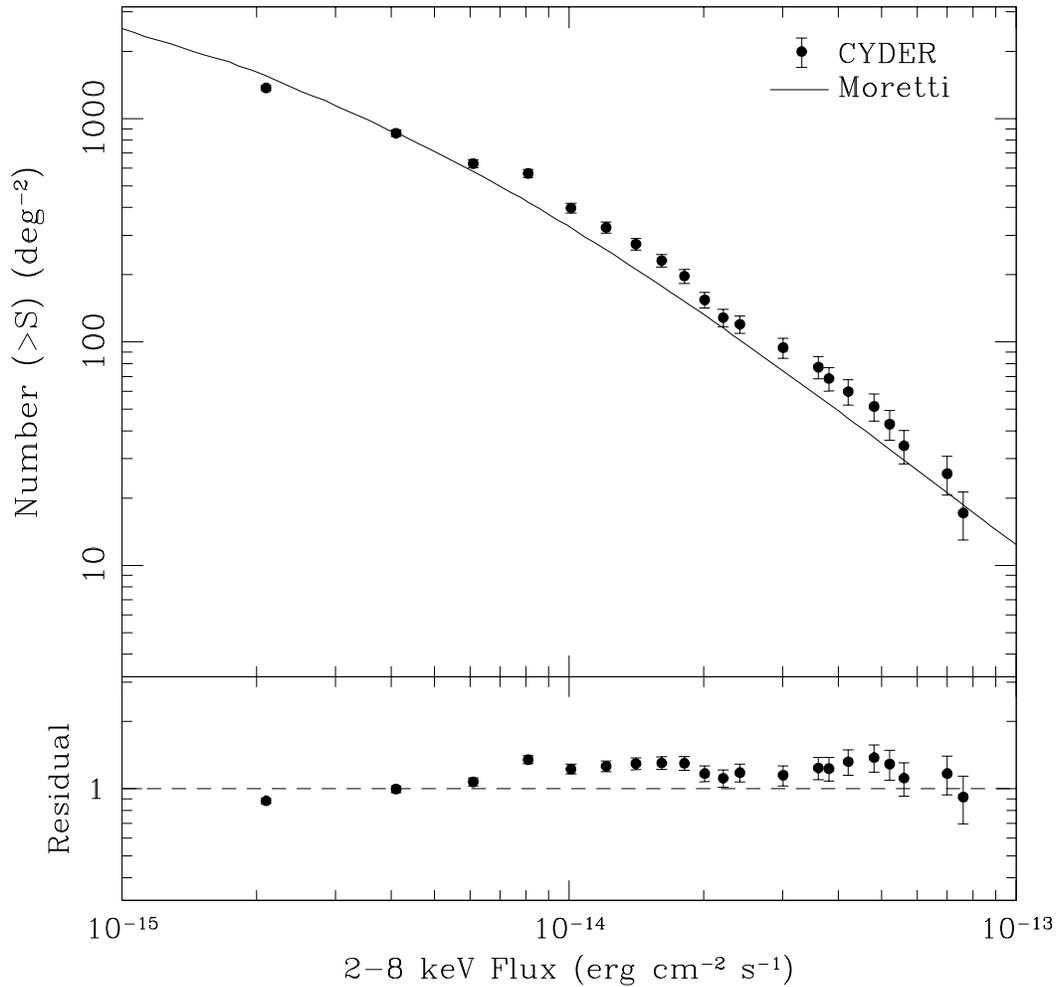}
\caption{Cumulative $\log N-\log S$ plot for sources in the CYDER survey detected 
in the hard X-ray band. In the upper panel, the solid line
shows the relation observed by \citet{moretti03} using a
combination of shallow and wide and deep pencil beam X-ray
surveys, while in the lower panel, residuals calculated as
the ratio of observed sources to the numbers obtained by
\citet{moretti03} are presented.}
\label{logn_cum}
\end{figure}

The cumulative flux distribution was calculated using
\begin{equation}
N(>S)=\sum_{i;S_i>S}\frac{1}{A_i(S_i)}
\label{cum_logn}
\end{equation}
\noindent
where $S_i$ is the observed hard X-ray flux of the $i$-th
source and $A_i$ is the maximum area over which that source
could be detected. The resulting $\log N-\log S$ relation
for the CYDER sample is shown in Figure~\ref{logn_cum}. This
curve is consistent with the relation computed by other
authors (e.g., \citealp{moretti03,ueda03}). In the lower
panel of Fig~\ref{logn_cum} we show the residuals after
comparing with the relation computed by \citet{moretti03}
using a combination of observational data in both shallow
wide-field and deep pencil beam X-ray surveys, showing that
the agreement is good.

\begin{figure}[!ht]
\figurenum{3}
\includegraphics[angle=0,scale=0.7]{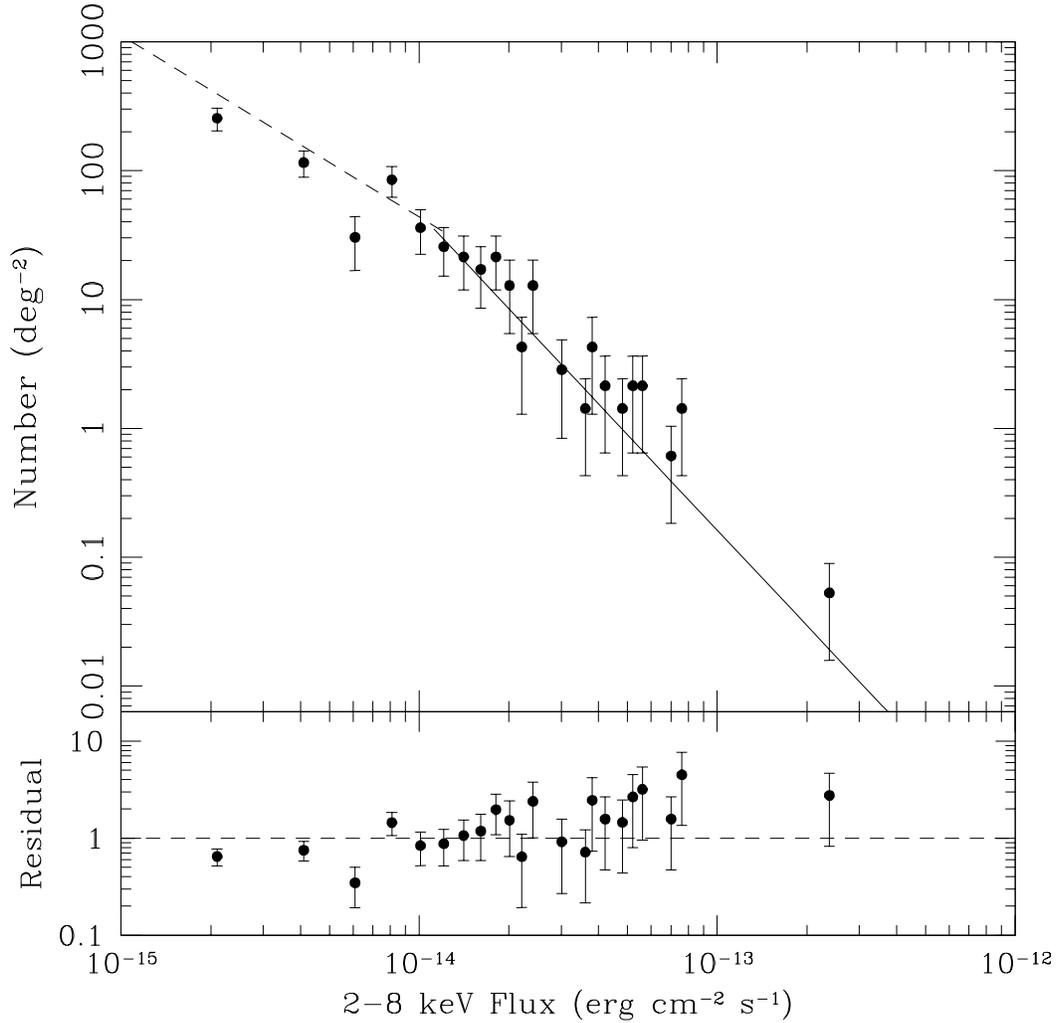}
\caption{Differential $\log N-\log S$ plot. Upper panel shows CYDER data, while 
the solid line shows the best-fit to the SEXSI counts. In
the lower panel, the residuals computed as the ratio between
the best-fit curve and the data are shown. Even tough some
scatter is present, the fit provides a good description of
the distribution of CYDER sources, with a reduced $\chi^2$
of 1.37.}
\label{logn}
\end{figure}

One significant problem with the cumulative $\log N-\log S$
relation is that the errors in each bin are not
independent. The differential flux distribution can be used
to avoid this problem. This relation can be expressed as:
\begin{equation}
n(S)_i=\sum \frac{N_i}{\Delta S_i A_i (S_i)}
\label{diff_logn}
\end{equation}
\noindent
where in this case the sample was binned with a bin size of
$2\times 10^{-15}$~erg~cm$^{-2}$~s$^{-1}$. $N_i$ is the
number of sources in the i-th bin, $\Delta S_i$ is the size
of the bin and $A_i$ is the the total area over which
sources in this bin could be detected. In Figure~\ref{logn}
the resulting differential $\log N-\log S$ is shown. These
results were compared to the relation reported by
\citet{harrison03}, which fitted the SEXSI data with a
broken power-law given by
\begin{equation}
n(S)=(46.8\pm 2.1)\left(\frac{S}{10^{-14}}\right)^{-2.46\pm 0.08}
\end{equation}
\noindent
for $S>1.25\times 10^{-14}$~ergs~cm$^{-2}$~s$^{-1}$ and
\begin{equation}
n(S)=(43.65\pm 2)\left(\frac{S}{10^{-14}}\right)^{-1.41\pm 0.17}
\end{equation}
\noindent
for fainter sources. As it is shown in the lower panel of
Figure~\ref{logn}, this parametrization provides a good fit
to the CYDER data, even though some scatter is present. A
$\chi^2$ test to this fit compared to the observed data gave
a reduced $\chi^2$ of 1.37.

\subsection{Optical}

\begin{figure}[!ht] 
\figurenum{4} 
\includegraphics[angle=270,scale=0.7]{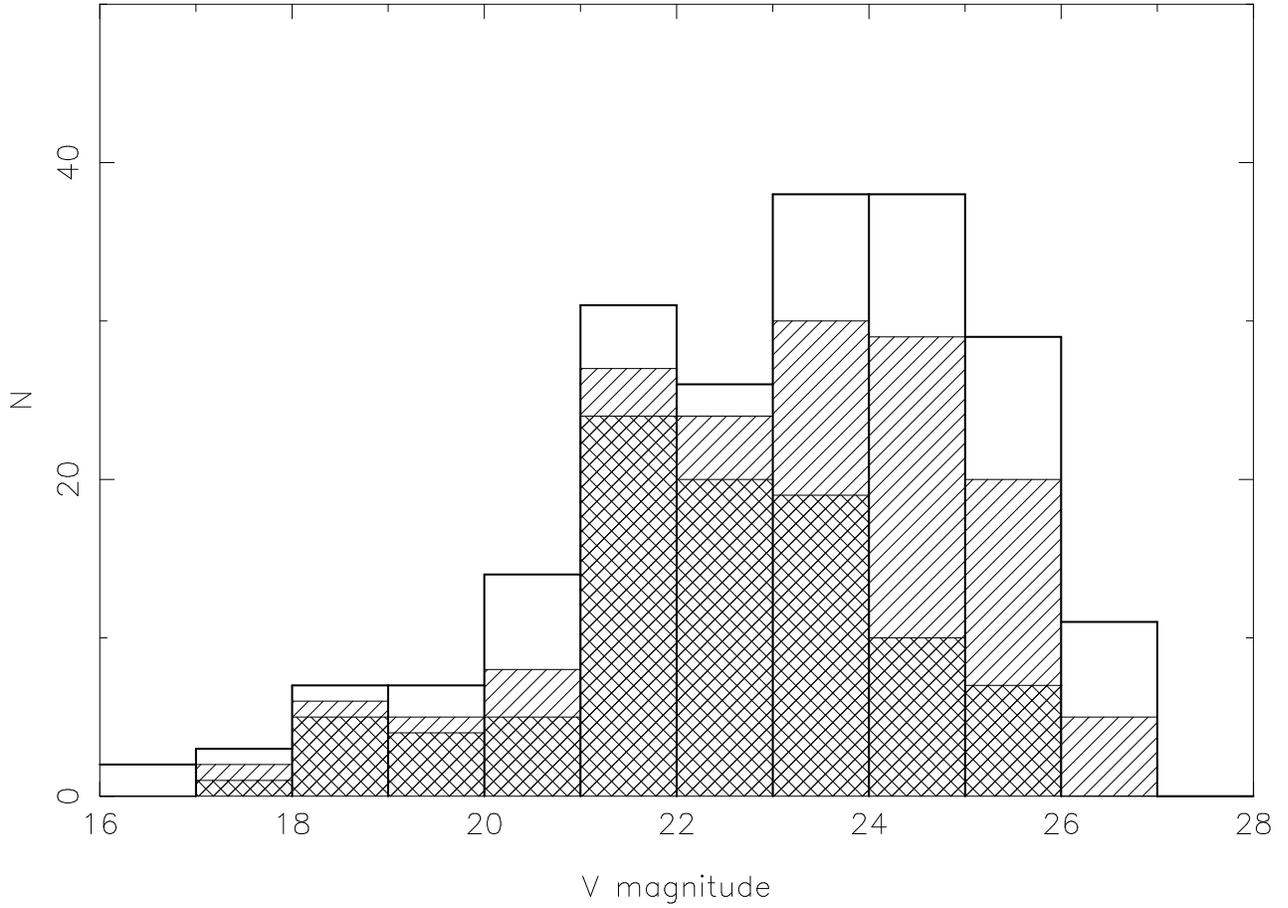} 
\caption{$V$ magnitude distribution for X-ray sources with
detected optical counterparts in the CYDER fields. Hatched
histogram shows the magnitude distribution for sources
targeted for optical spectroscopy, while the cross-hatched
histogram shows the distribution of sources successfully
identified. While sources were selected for spectroscopy
independent of their optical properties, it is clear that
spectroscopic identifications are much more efficient for
X-ray sources brighter in the optical.}
\label{v_dist}
\end{figure}

\begin{figure}[!ht]
\figurenum{5} 
\includegraphics[angle=270,scale=0.7]{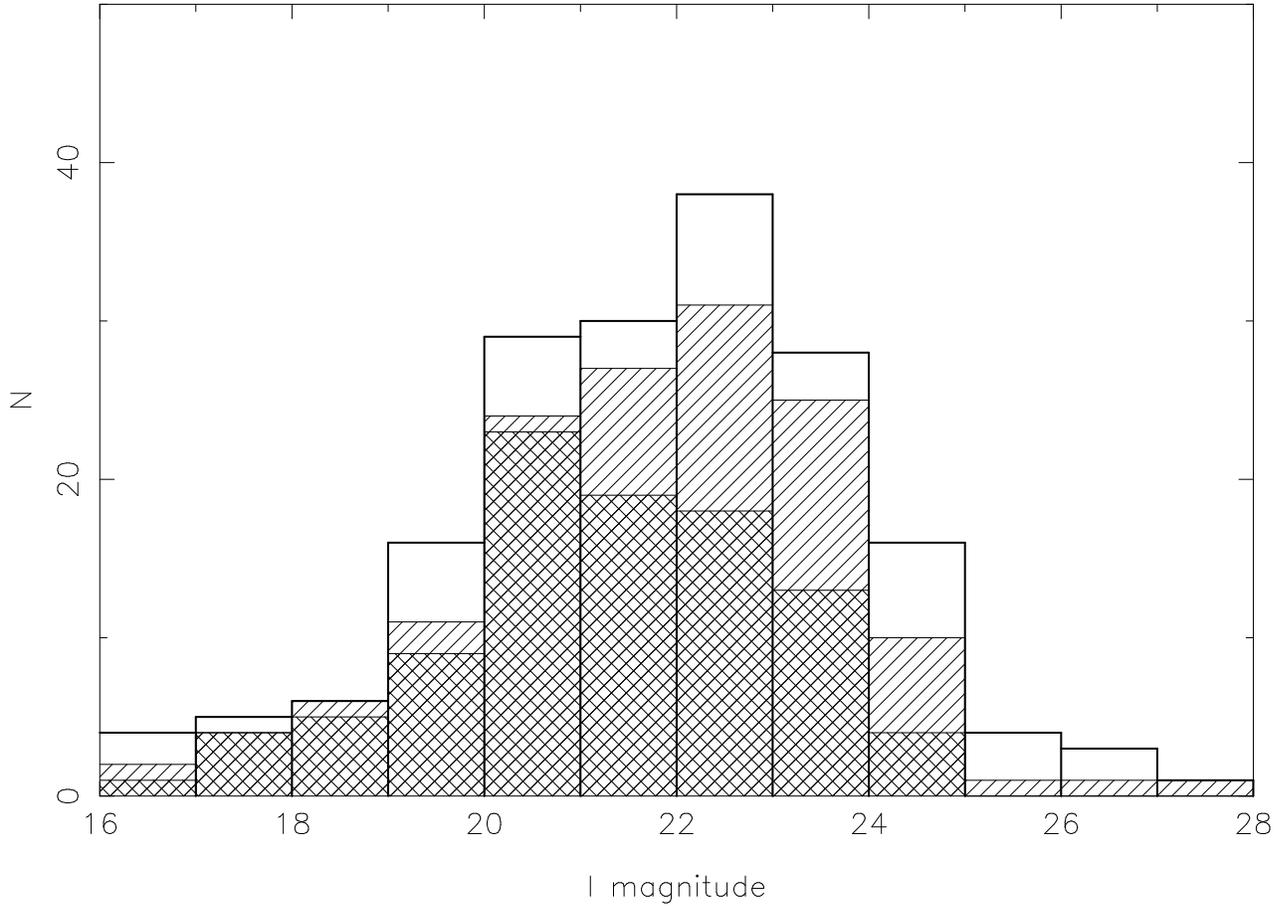} 
\caption{$I$ magnitude distribution for X-ray sources with detected $I$-band 
counterparts. The magnitude distribution for sources
targeted for spectroscopy is shown by the hatched histogram,
while the distribution for sources successfully identified
is shown by the cross-hatched histogram. Again, the
efficiency of spectroscopic identifications is much higher
for sources brighter in the optical bands.}
\label{i_dist}
\end{figure}

\begin{figure}[!ht] 
\figurenum{6} 
\includegraphics[angle=270,scale=0.7]{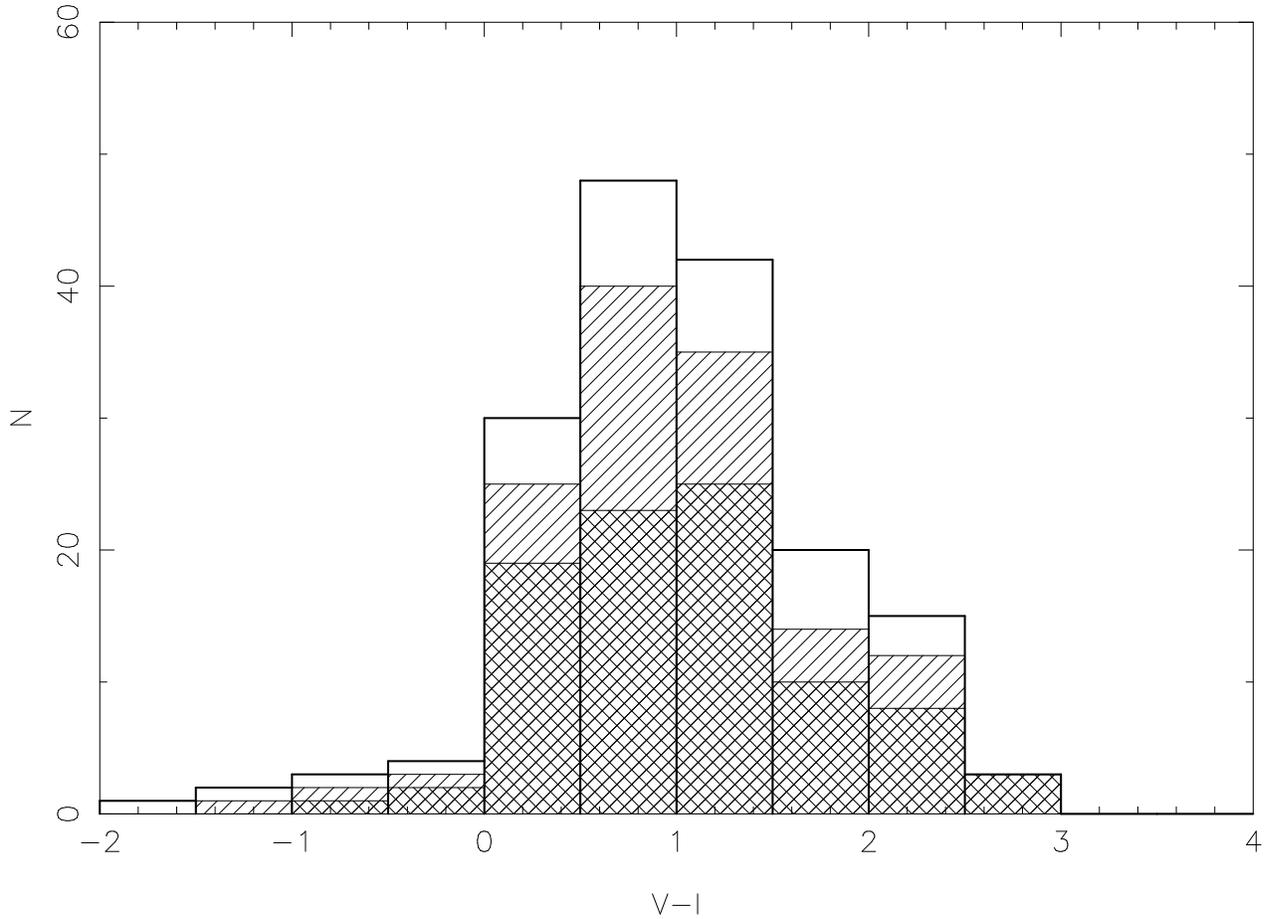} 
\caption{$V-I$ color distribution for X-ray sources detected in the optical bands.
Hatched histogram shows the distribution for sources targeted for
spectroscopy while the cross-hatched histograms shows the distribution
for sources successfully identified. In this case, these distributions
are very similar, with a K-S confidence level for the null hypothesis
of 99.89\%, showing that the efficiency of spectroscopic
identifications is independent of the $V-I$ color of the optical
counterpart.}
\label{vi_dist}
\end{figure}

207 of the total 267 X-ray sources (77\%) were detected in
our optical $V$-band images, searching in a radius of
2.5$''$ ($\sim$ 4 times the typical seeing) around the
centroid of the X-ray emission. The average offset between
an X-ray source and the nearest optical counterpart is $\sim
1.3''$ with a standard deviation of $\sim 0.5''$. In our
optical images we detected typically $\sim 60,000$ sources
in $30'\times 30'$ so the chance of having a random source
in a 2.5$''$ radius is $\sim 36$\% and therefore our choice
of 2.5$''$ as the maximum allowed offset is reasonable to
avoid spurious associations. In cases where more than one
counterpart was found inside this radius the closest optical
source to the X-ray centroid was assumed to be the right
counterpart. The $V$ magnitude distribution for the X-ray
sources with detected optical counterparts in the CYDER
fields is shown on Figure \ref{v_dist}. X-ray sources cover
the range from $V\simeq 16$ to 26 mags and higher (fainter
than our optical magnitude limit). The hatched histogram in
Figure~\ref{v_dist} shows the magnitude distribution of
sources targeted for spectroscopy, while the cross-hatched
histogram shows the distribution for sources with
spectroscopic identifications. A K-S test performed
comparing the total $V$ magnitude distribution to the
magnitude distribution for sources targeted for spectroscopy
revealed that the hypothesis that both distributions are
drawn from the same parent distribution (the null
hypothesis) is accepted at the 98.7\% confidence
level. However, the effect of the optical flux in the
efficiency of spectroscopic identifications can be seen by
comparing the magnitude distribution for sources
successfully identified and the total sample, namely the
incompleteness of the sample with spectroscopic
identification at the faint optical flux end is evident in
this figure, even though the target selection was
independent of the optical properties of the source. This
effect is also observed in the $I$-band
(Figure~\ref{i_dist}), where 181 X-ray sources counterparts
were detected (68\%), a lower number than in the deeper
$V$-band images. In this case, the decrease in the
efficiency of spectroscopic identifications with decreasing
optical flux is also evident in Figure~\ref{i_dist}. The
average $V-I$ color for X-ray sources with optical
counterparts is $0.92$; its distribution, shown in
Figure~\ref{vi_dist}, shows that the efficiency of the
spectroscopic identifications is independent of the $V-I$
color of the optical counterpart. A K-S test performed
comparing the sample with spectroscopic identification to
the total sample show that both distributions are drawn from
the same parent distribution with a confidence level of
99.89\%.

53 out of 267 ($\sim 20$\%) sources were not detected in any of the
two optical bands. The vast majority of these sources are also very
faint in X-rays, so that many of them were only detected in the soft
band, that is more sensitive in Chandra. This does not imply that they
have an intrinsically soft spectrum. In fact, given the known relation
between hardness ratio and X-ray flux \citep{giacconi01}, it is
plausible that these sources are hard and therefore good candidates to
be obscured AGN. 5 of these sources were only detected in the hard
band, and therefore should have a very hard spectrum in X-rays, that
combined with the fact that are very faint in the optical bands makes
them good candidates to be obscured AGN at relatively high
redshifts. This lack of detection of X-ray faint sources in optical
images acts as a selection effect against the study of obscured
AGN. However, this bias can be overcome by studying these sources in
the near infrared \citep{gandhi04}, where the effects of dust
obscuration are much smaller. In a following paper (F. Castander, in
prep) properties of these sources in the near infrared bands will be
presented.

\begin{figure}[!ht]
\figurenum{7}
\includegraphics[angle=270,scale=0.7]{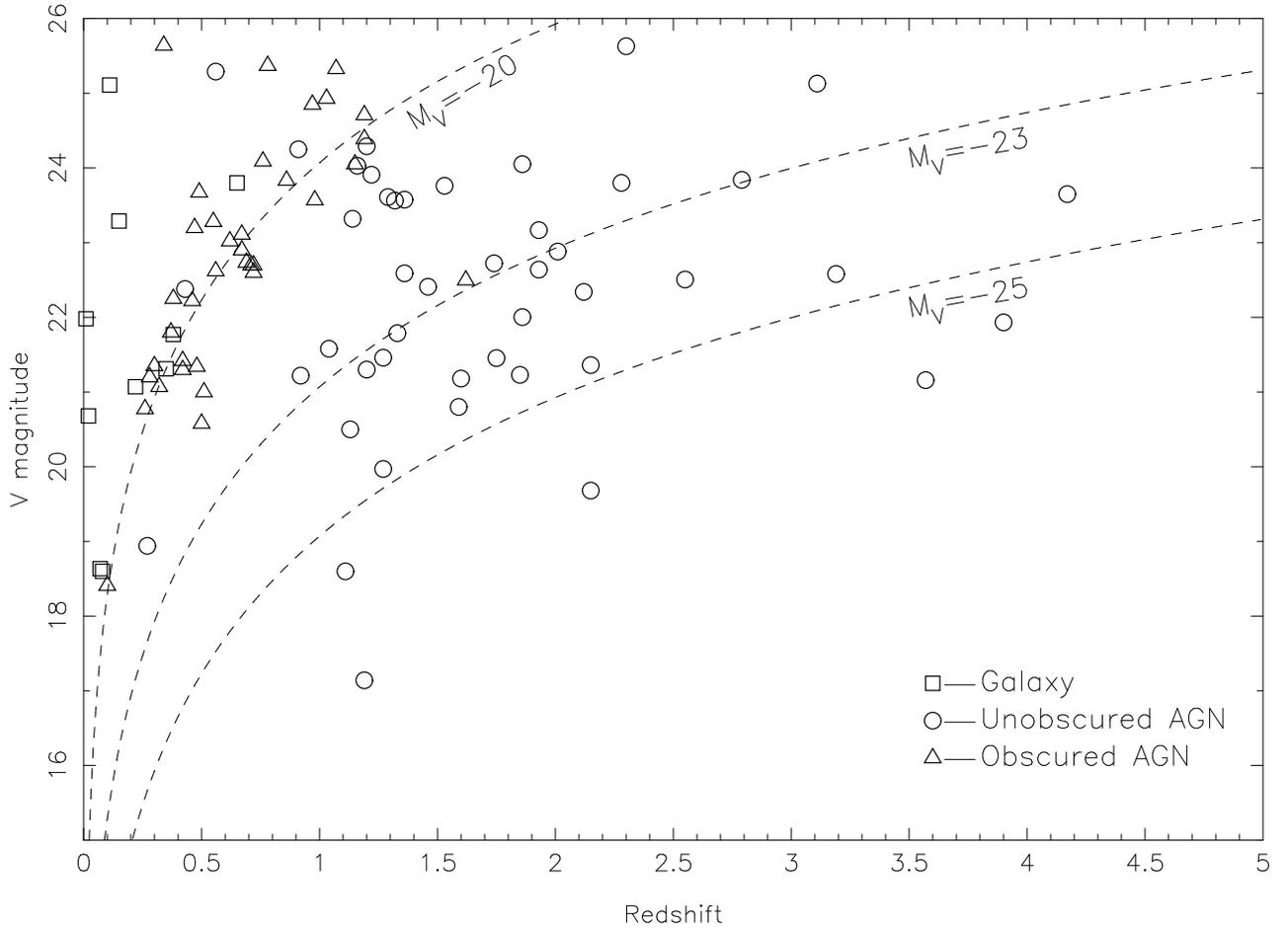}
\caption{$V$-band magnitude versus redshift for sources with spectroscopic 
identification. {\it Circles}: unobscured (Type 1; broad lines)
AGN. {\it Triangles}: obscured (Type 2; narrow lines) AGN. {\it
Squares}: Galaxies. As expected, most of the high redshift sources are
broad line AGN with optical magnitudes in the 22-24 mag. range. Dashed
lines shows the position of constant absolute magnitude $M_V=-20,-23$
and -25. While almost all the sources classified as normal galaxies or
obscured AGN have $M_V\ga -21$, unobscured (broad lines) AGN have
$M_V<-22$.}
\label{v_z}
\end{figure}

The $V$-band magnitude versus redshift plot
(Figure~\ref{v_z}) reveals how the source composition
changes with redshift, and how it is potentially affected by
the implicit optical flux cut for spectroscopy. While at low
redshift ($z<1$) we find mostly obscured AGN and normal
galaxies, characterized by an absolute optical magnitude
$M_V\ga -21$, at higher redshift they become too faint in
the optical bands and therefore only unobscured AGN (that
have mostly $M_V<-22$) can be found. This implies that our
survey may be biased against detecting obscured AGN at high
redshift. This effect is investigated in more detail in
\S~\ref{disc}.

\subsection{Correlations}

Sources with spectroscopic identification were classified
using a combination of optical and X-ray criteria. X-ray
sources showing stellar spectra were classified as
stars. For extragalactic sources, the X-ray luminosity was
computed from the observed X-ray flux using the relation
\begin{equation}
L_X=4\pi d_L^2 f_X
\end{equation}
\noindent
where $d_L$ is the luminosity distance calculated for the
assumed cosmology. This luminosity is therefore the
uncorrected, observed frame X-ray luminosity. No attempt was
made to correct for dust obscuration or k-corrections given
that for most sources the number of observed counts was too
small to perform spectral fitting and therefore to calculate
the neutral hydrogen column density $N_H$ or the intrinsic
spectral shape.

In order to separate X-ray emission generated by AGN
activity from the emission coming from X-ray binaries and
star formation in galaxies we used a simple X-ray luminosity
threshold criterion. Locally, the most X-ray luminous star
forming galaxy known (NGC 3256) has a total X-ray luminosity
$L_X\simeq 8\times 10^{41}$~ergs~s$^{-1}$ in the (0.5-10)
keV band \citep{lira02}. Another source of luminous X-ray
emission is the presence of hot gas in elliptical galaxies,
which at low redshift is extended and therefore easily
separated from AGN emission; at high redshift it is not
resolved and thus harder to separate from AGN
activity. However, according to the \citet{osullivan01}
catalog of elliptical galaxies with detected X-ray emission,
only a few normal galaxies have
$L_X>10^{42}$~ergs~s$^{-1}$. Therefore, we adopted
$L_X=10^{42}$~ergs~s$^{-1}$ in the total (0.5-8 keV) band as
the threshold separating sources dominated by AGN activity
from those dominated by star formation or other processes in
a galaxy. Given the relatively low number of galaxies found
in the survey, we expect this classification method to have
a small effect on the total numbers of AGN reported.

Objects with a total X-ray luminosity
$L_X<10^{42}$~ergs~s$^{-1}$ and narrow emission or
absorption lines (velocity dispersion less than
1000~km~s$^{-1}$) were classified as normal galaxies, while
sources with $L_X>10^{42}$~ergs~s$^{-1}$ were classified as
unobscured (type 1) or obscured (type 2) AGN depending on
whether they show broad or narrow lines on their optical
spectrum. Furthermore, sources with
$L_X>10^{44}$~ergs~s$^{-1}$ are called QSO-1, or simply QSO,
if they have broad lines or QSO-2 if the lines are narrow,
but they are still considered AGN.

\begin{figure}[!ht]
\figurenum{8}
\includegraphics[angle=270,scale=0.7]{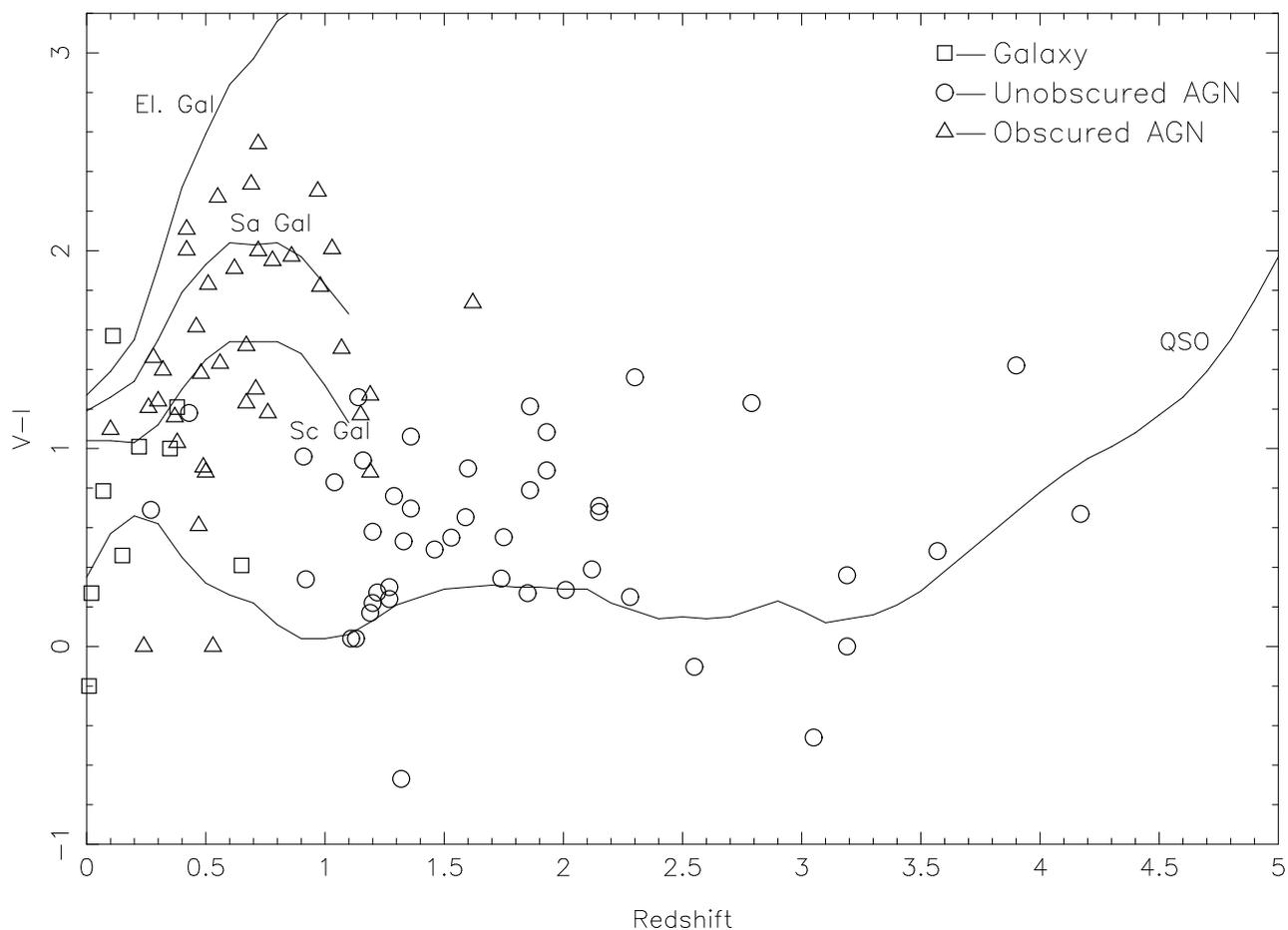}
\caption{$V-I$ versus redshift for X-ray sources with spectroscopic
 identifications. {\it Circles}: unobscured AGN. {\it
 Triangles}: obscured AGN. {\it Squares}:
 Galaxies. Continuous lines show the expected color as a
 function of redshift for each type of source. Synthetic
 colors for galaxies were computed using the \citet{fioc97}
 galaxy spectrum models while for the QSO track the Sloan
 Digital Sky Survey composite quasar spectrum
 \citep{vandenberk01} was used.}
\label{vi_z}
\end{figure}

For sources with spectroscopic identification,
Figure~\ref{vi_z} shows the $V-I$ color as a function of
redshift. X-ray sources identified as Type 1 AGN (broad
emission lines) fall near the position expected for QSOs,
calculated convolving the optical filters with the Sloan
Digital Survey Composite Quasar Spectrum
\citep{vandenberk01}. Galaxies/Type 2 AGN, which are only
detected up to $z\sim 1.5$, are located in the region
expected for galaxies ranging from Elliptical to Sb types
and have redder colors than Type 1 AGN. The expected colors
for each type of galaxy at a given redshift were calculated
using the galaxy spectrum models of \citet{fioc97} assuming
that there is no evolution in the spectrum with
redshift. From Figure~\ref{vi_z} it is clear that objects
classified as obscured AGN have redder colors, consistent
with those of the host-galaxies. In fact, obscured AGN have
an average $V-I$ color of 1.46 with a standard deviation of
0.58, while unobscured AGN have an average color of 0.56 and
standard deviation of 0.46

\begin{figure}[!ht]
\figurenum{9}
\includegraphics[angle=270,scale=0.7]{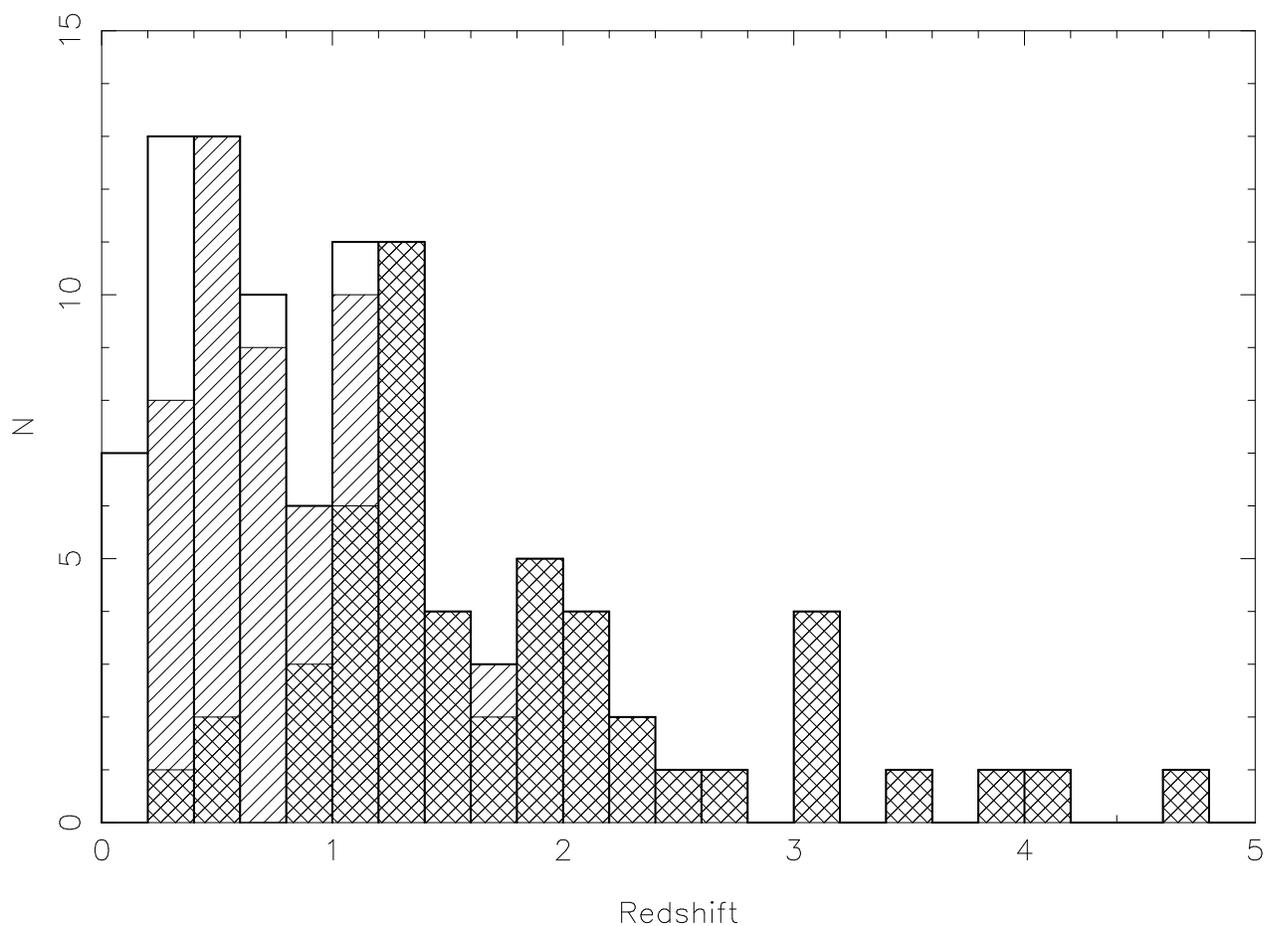}
\caption{{\it Solid line}: Redshift distribution of extragalactic 
X-ray sources in the CYDER survey. Stars were removed for
clarity. {\it Cross-hatched histogram}: redshift distribution for
unobscured (broad lines) AGN, that dominate the population at high
redshift ($z>1.3$). {\it Hatched histogram}: distribution for sources
with $L_X>10^{42}$~ergs~s$^{-1}$ (i.e., AGN dominated). In this case
the broad peak of the distribution is found at $z\sim 1$.}
\label{red_dist}
\end{figure}

The redshift distribution for sources with spectroscopic
identification is shown in Figure~\ref{red_dist}. When the
whole sample of X-ray sources is considered, this
distribution has a maximum at very low redshift, $z\simeq
0-0.6$. However, when only sources with
$L_X>10^{42}$~ergs~s$^{-1}$ (i.e., those dominated by AGN
activity) are included the peak is displaced to higher
redshifts, namely $z\sim 1$. As it is shown by the hatched
distribution in Figure \ref{red_dist}, the high redshift
population ($z>1.3$) is completely dominated by broad line
AGN (most of them quasars with
$L_X>10^{44}$~ergs~s$^{-1}$). This is explained by the high
optical luminosity of these objects, which makes them easier
to identify, even at large distances and by the lack of near
infrared information at this point, that is very useful to
detect obscured AGN, in particular at high redshift
\citep{gandhi04}.

\begin{figure}
\figurenum{10}
\includegraphics[angle=270,scale=0.7]{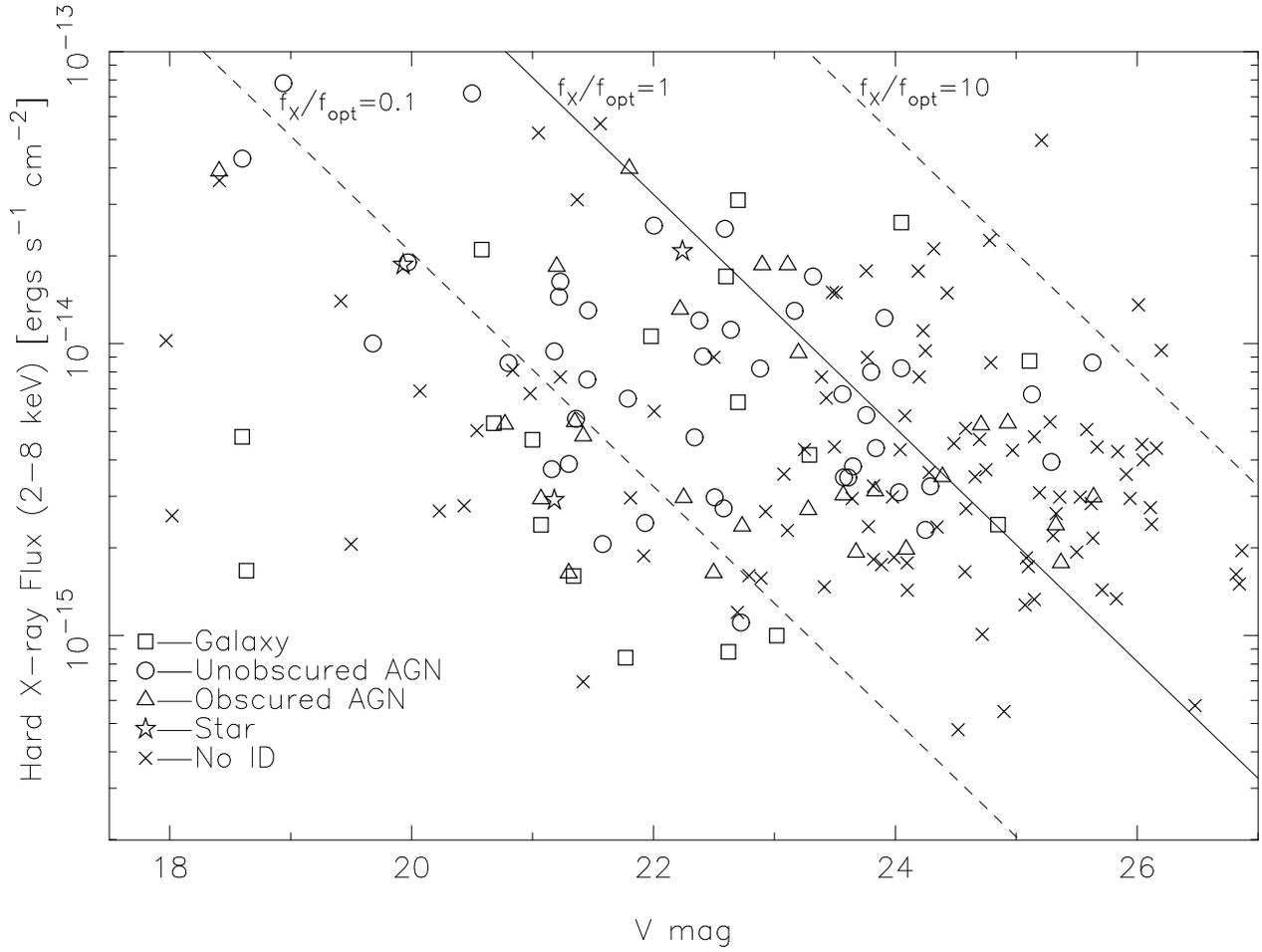}
\caption{Hard (2-8 keV) X-ray flux vs $V$ band magnitude. 
{\it Circles}: unobscured AGN. {\it Triangles}: obscured AGN. {\it
Squares}: Galaxies. {\it Five pointed Stars}: Stars with X-ray
emission. {\it Crosses}: sources without spectroscopic
identification. The solid line shows the locus of sources with $\log
f_X/f_{\rm opt}=0$, while dashed lines show the position of sources
with $\log f_X/f_{\rm opt}=\pm 1$.}
\label{fx_v}
\end{figure}

In order to investigate possible relations between X-ray and
optical emission for different classes of sources, in
Figure~\ref{fx_v} we plot hard X-ray flux versus $V$-band
magnitude. Most of the sources are located in the region
bounded by $\log f_X/f_{\rm opt}=\pm 1$. Starburst galaxies
detected in X-rays are typically bright in the optical bands
and faint in X-rays, and are therefore characterized by
$\log f_X/f_{\rm opt}<-1$ (see \citet{hornschemeier03} and
references therein). Unobscured (type 1) AGN/Quasars are
located around the $\log f_X/f_{\rm opt}=1$
\citep{giacconi02} position although the scatter is large,
while obscured (type 2) AGN/Quasars have in general $\log
f_X/f_{\rm opt}>1$ since most of the optical light from the
central engine is blocked from our view but low-luminosity
examples of obscured AGN can be found also with $\log
f_X/f_{\rm opt}\simeq 0$ as it will be discussed in
\S5. Unidentified sources at high $f_X/f_{\rm opt}$ are
unlikely to be unobscured AGN because their broad emission
lines would have been easy to see in the optical spectra, so
they are probably obscured AGN.

\begin{figure}
\figurenum{11}
\includegraphics[angle=270,scale=0.7]{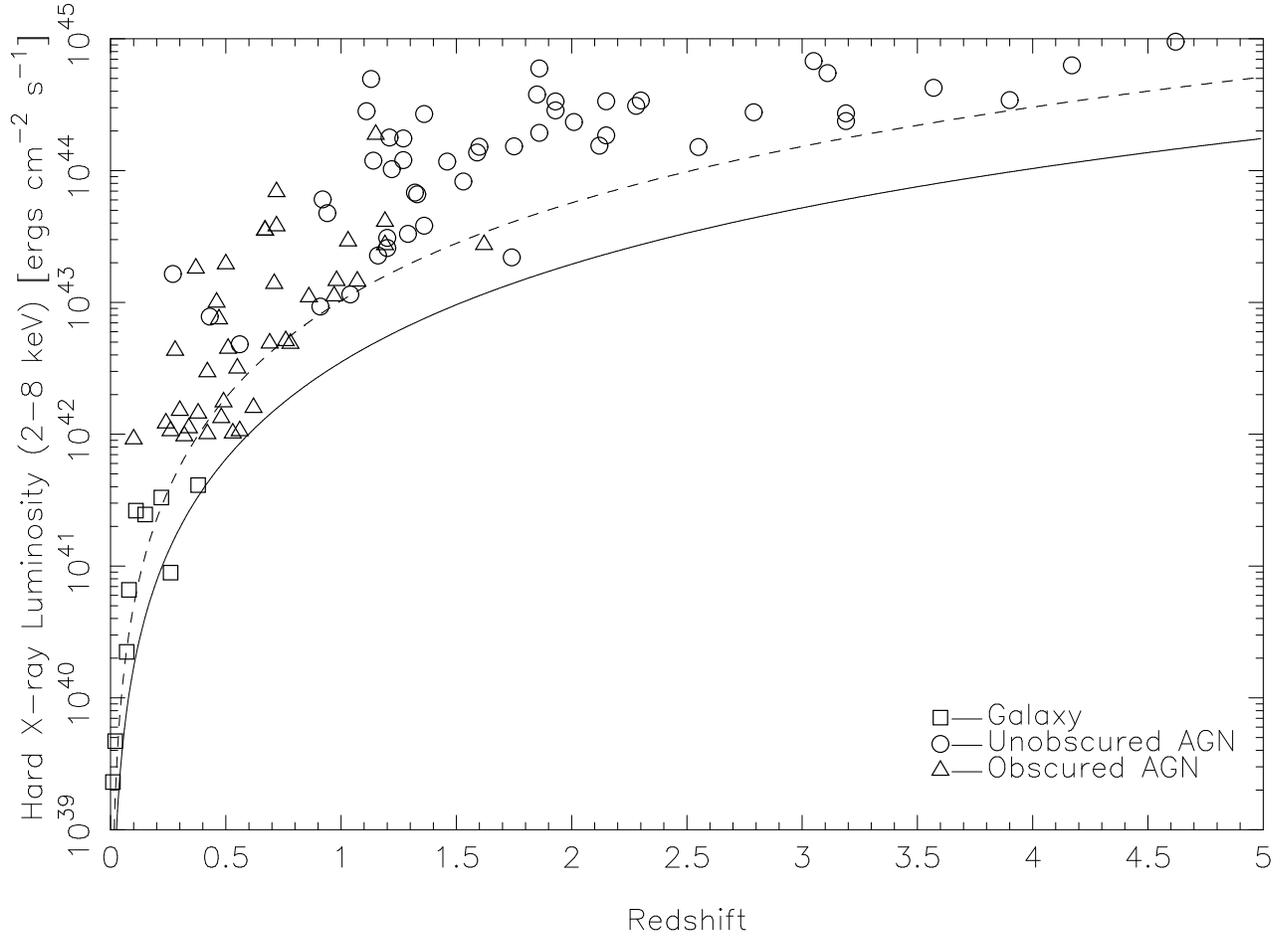}
\caption{Hard (2-8 keV) X-ray luminosity versus redshift for 
sources with spectroscopic identification. Symbols are the
same as in Figure \ref{vi_z}. {\it Solid line}: flux limit
for a simulated observation of a source detected with 5
counts in the total (0.5-8) keV band in 60 ks with ACIS-I on
board Chandra. {\it Dashed line}: X-ray luminosity for a
source with optical magnitude $V=25.0$ mags and $f_X/f_{\rm
opt}=1$.}
\label{l_z}
\end{figure}

Figure \ref{l_z} shows the observed-frame hard X-ray (2-8
keV) luminosity versus redshift diagram for sources with
spectroscopic redshifts. If we assume a flux limit of
$6.9\times 10^{-16}$~ergs~cm$^{-2}$~s$^{-1}$ (which would
yield a total of 5 counts in the 0.5-8 keV band in a
$Chandra$ ACIS-I 60 ks observation for $\Gamma=1.7$), the
solid line in Figure \ref{l_z} shows the detection limit for
X-ray sources in our survey. If an optical magnitude of
$V=25$ mag is taken as the approximate flux limit for
spectroscopy (there are fainter sources for which
spectroscopy is possible, but the identification relies in
the presence of strong emission lines) and a ratio of X-ray
to optical emission of $f_X/f_{\rm opt}=1$ is assumed, then
the dashed line in Figure \ref{l_z} shows our limiting
magnitude as a function of redshift for sources with
spectroscopy. This explains why incompleteness of the
spectroscopic sample is particularly important at high
redshift, where the fraction of X-ray sources with
spectroscopic identification declines.

If the same material that is causing the absorption of
X-rays is responsible for the extinction in the optical
bands, then a relation between these two quantities can be
expected, namely the reddest sources in the optical (higher
value of $V-I$) should also be the X-ray hardest sources. A
typical way to quantify the steepness of the X-ray spectrum
is using the hardness ratio (HR) defined as
\begin{equation}
HR=\frac{H-S}{H+S}
\end{equation}
\noindent
where $H$ and $S$ are the count rates in the hard and soft
X-ray bands respectively.

\begin{figure} 
\figurenum{12} 
\includegraphics[angle=270,scale=0.7]{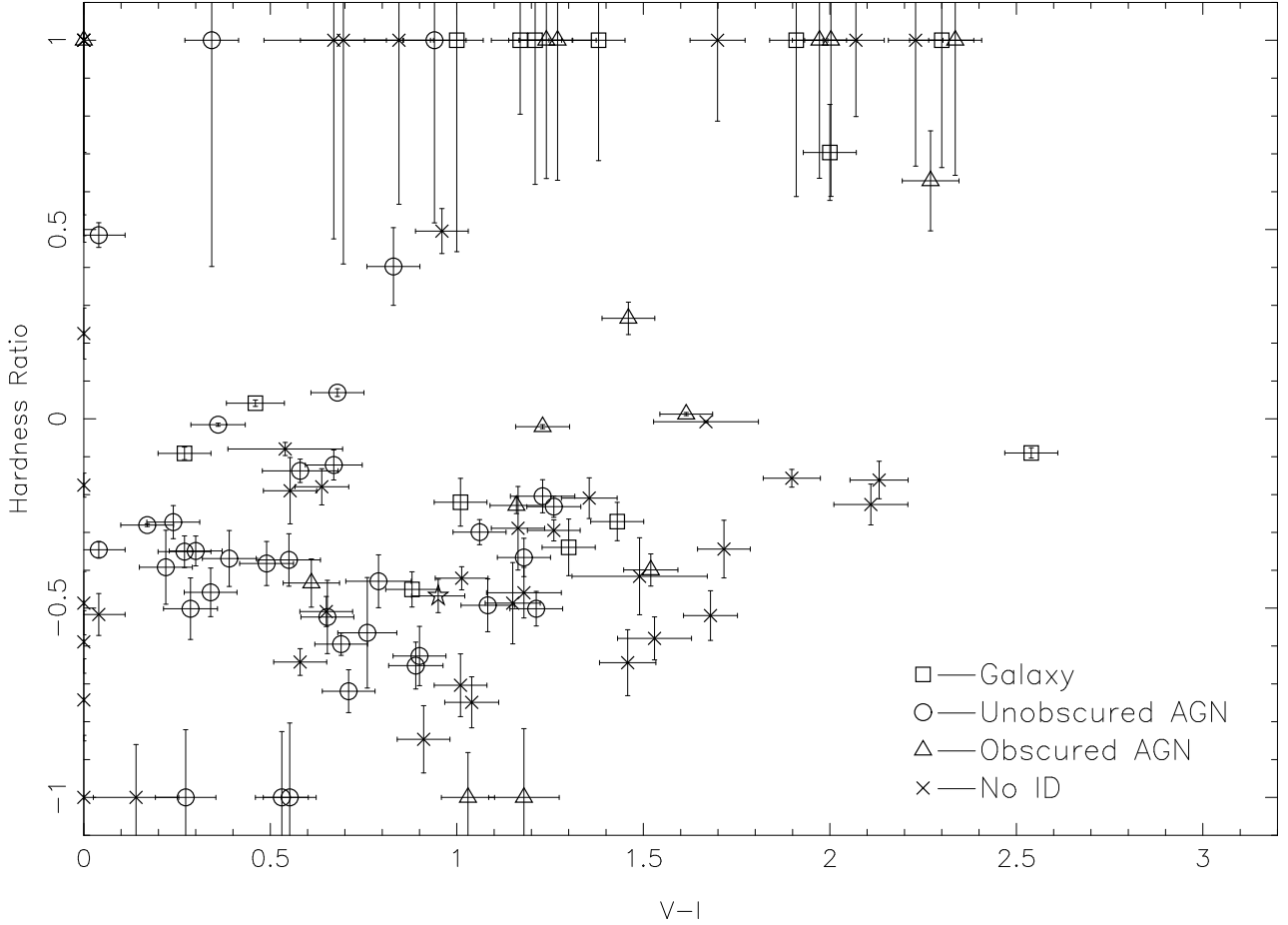} 
\caption{HR, defined here as (H-S)/(H+S) where H and S 
are the hard and soft X-ray band counts respectively, versus
$V-I$ color; a source with HR=1 was only detected in the
hard band, while one with HR=-1 was only detected in the
soft band. Sources with fewer than 50 counts observed in the
soft band and not detected in the hard band are not show in
this plot. Symbols are the same as in Figure
\ref{fx_v}. Given the spread in intrinsic $V-I$ color and
X-ray spectral shape, a clear correlation between HR and
optical color is not observed, however a general trend can
be seen in the sense that objects classified as type 2 AGN
are redder and have larger HR values, consistent with the
presence of obscuration affecting both X-ray and optical
emission.}
\label{hr_vi}
\end{figure}

In Figure~\ref{hr_vi} the HR versus $V-I$ optical color is
presented. In this diagram, no clear relation between HR and
optical color is observed. The absence of a correlation can
be explained by the differences in the intrinsic optical and
X-ray spectrum for different types of sources detected in
the X-ray bands, independent of the amount of obscuration
present. Note, however, that in general sources optically
classified as obscured AGN are redder (larger $V-I$ colors)
than unobscured AGN as was previously observed on
Figure~\ref{vi_z} and also tend to have higher values of
HR. This lack of a strong relationship between HR and
optical color even for sources classified as AGN-dominated
can be explained in part by the effects of K-corrections
caused by the different redshifts of the sources and by
changes in the intrinsic spectrum with parameters other than
obscuration, e.g. luminosity \citep{ho99}. Sources without
spectroscopic identification (crosses on Figure \ref{hr_vi})
have in general redder colors than unobscured AGN, that are
similar to the colors of spectroscopically confirmed
obscured AGN and therefore are consistent with being
moderately obscured AGN at relatively high redshift ($z\ga
1$). Most of these sources however present a soft X-ray
spectrum, that can be explained if these sources are at
moderately high redshift so that the observed frame Chandra
bands traces higher energy emission, that is less affected
by absorption. However, this is highly speculative, and the
final answer about the nature of these optically faint X-ray
sources will come from either deeper optical spectroscopy or
from the near infrared data.

\subsection{Identifications}

Of the 267 X-ray sources detected in the CYDER fields, 106 were
identified using optical spectroscopy. While the fraction of sources
identified is biased toward higher optical fluxes (see Figures
\ref{v_dist} and \ref{i_dist}), Figure~\ref{vi_dist} shows that the
optical colors of the sources in the sample both targeted and
identified by optical spectroscopy follow a similar distribution as
those of the total sample.

The redshift distribution of the sample with spectroscopic
identification is presented in Figure~\ref{red_dist}. X-ray
sources in this sample span a wide range in redshift,
$0<z<4.6$. The mean redshift for our extragalactic sample
with spectroscopic identification is $<z>=1.19$ and the peak
is located at a low redshift, $z\simeq 0.2-0.6$. When only
the sources dominated by AGN activity (i.e.,
$L_X>10^{42}$~ergs~s$^{-1}$) are considered, the mean
redshift is $<z>=1.34$ while the peak is at $z\simeq
0.5$. For sources optically classified as unobscured AGN,
the average redshift is $<z>=1.82$ and the peak is at
$z_p=1.3$. Therefore, we conclude that the nature of the
{\it identified} X-ray sources changes as a function of
redshift. At $z<0.3$, the sample is dominated by normal
galaxies ($\sim 60$\%) and obscured AGN. In the $0.3<z<1$
region, just a few normal galaxies are found and the
population is dominated by obscured AGN (77\%), while at
$z>1$ the vast majority of the sources found are unobscured
AGN.

\begin{figure}
\figurenum{13}
\includegraphics[angle=270,scale=0.7]{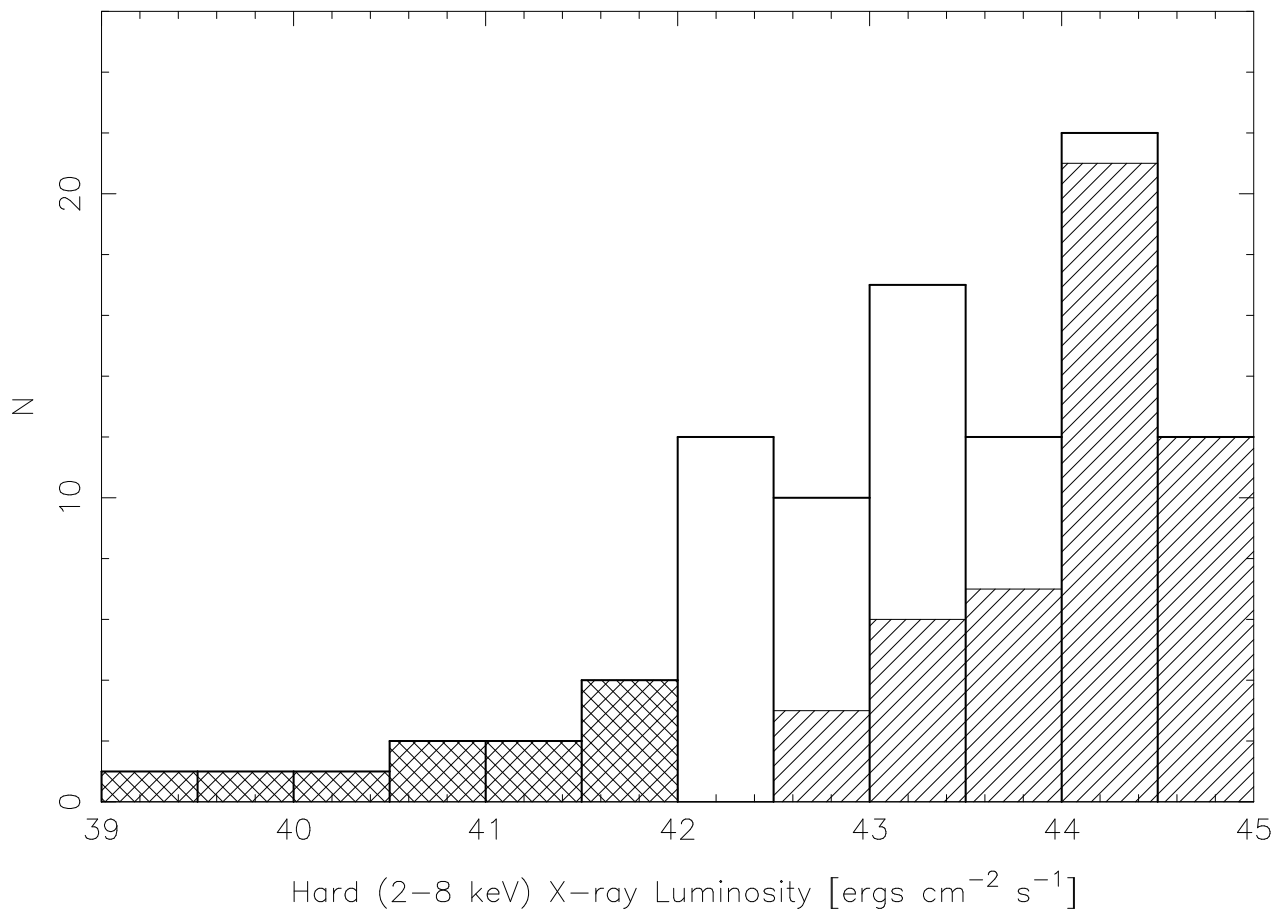}
\caption{Hard (2-8 keV) luminosity distribution for the 106 sources 
with spectroscopic identification. {\it Hatched histogram:}
luminosity distribution of unobscured AGN. {\it
Cross-hatched histogram}: luminosity distribution for
galaxies (i.e., $L_X<10^{42}$~ergs~s$^{-1}$). Unobscured AGN
dominate the higher luminosity part of the distribution,
while obscured AGN are the majority of the sources in the
$10^{42}<L<10^{44}$~ergs~cm$^{-2}$~s$^{-1}$ region.}
\label{l_dist}
\end{figure}

The hard X-ray luminosity distribution for the sample of
sources with spectroscopic redshift can be seen in
Figure~\ref{l_dist}. In terms of luminosity, the few sources
optically classified as galaxies detected in the X-ray
sample dominate the low luminosity bins. In the intermediate
X-ray luminosity bins
($10^{42}<L_X<10^{44}$~ergs~cm$^{-2}$~s$^{-1}$), most of the
sources are optically classified as obscured AGN, while in
the higher luminosity bins
($L_X>10^{44}$~ergs~cm$^{-2}$~s$^{-1}$) the vast majority of
the sources are optically identified as unobscured AGN. This
change of the source type as a function of X-ray luminosity
is further investigated in
\S~\ref{disc}.

\begin{figure}
\figurenum{14}
\includegraphics[angle=0,scale=0.7]{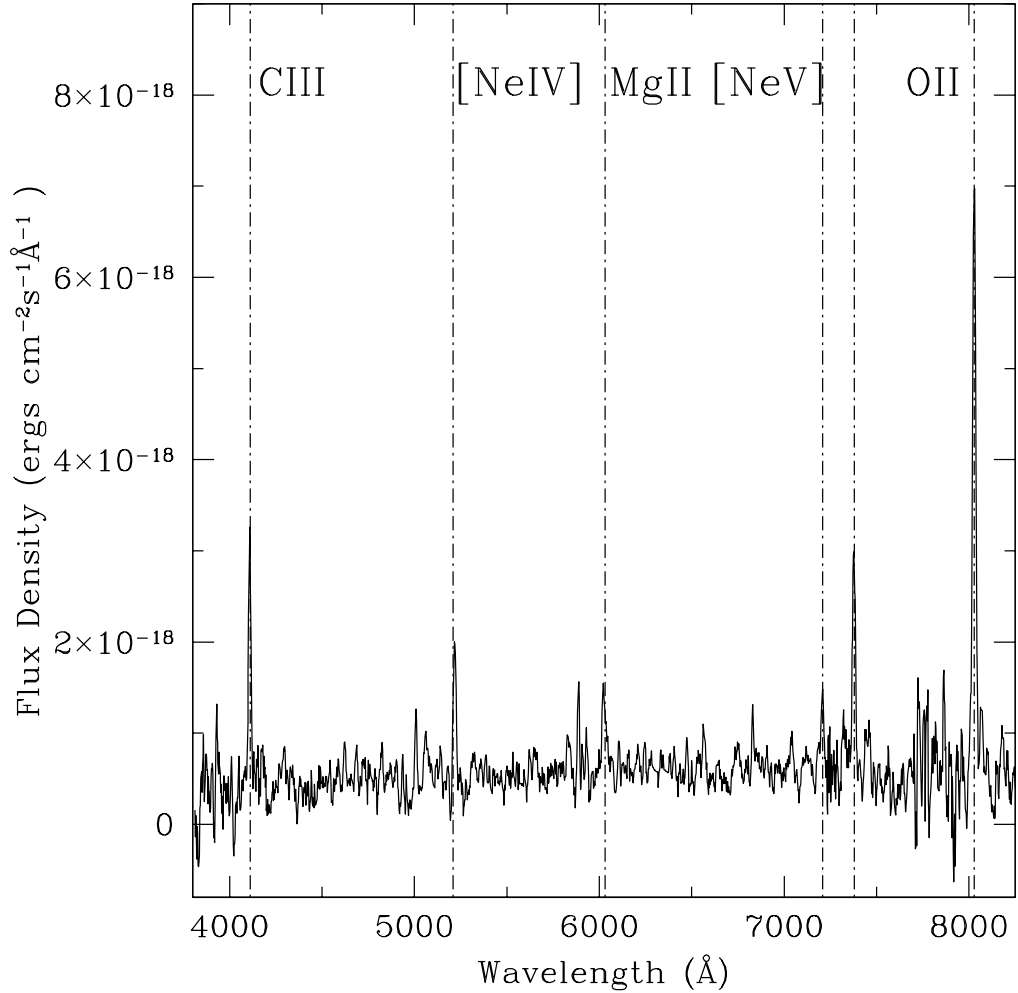}
\caption{Optical spectrum of CXOCY-J125315.2-091424, the only source 
classified as a type 2 quasar detected in the sample. Most significant
emission lines detected are identified, securing a redshift of
$z=1.154$ for this source. Narrow emission lines like CIII, MgII and
OII are clearly visible in the spectrum of this object.}
\label{qso2_spec}
\end{figure}

In our sample there is only one source classified as QSO-2 based on
its observed X-ray luminosity and optical spectrum:
CXOCY-J125315.2-091424 at $z=1.154$, located in the C2 field. 52
counts were detected in the hard X-ray band, while no emission was
detected in the soft X-ray band and therefore the hardness ratio is
-1. The optical spectrum of this source is presented in
Figure~\ref{qso2_spec}. Narrow emission lines like CIII, MgII and OII
are clearly visible in this spectrum and were used to calculate the
redshift of the source. The total (0.5-8 keV) observed X-ray
luminosity of this source is $L_X\simeq 2\times 10^{44}$~ergs~s$^{-1}$
making this the brightest obscured AGN in our sample. Given the
observed hardness ratio and redshift of this source, the expected
neutral hydrogen column density in the line of sight is $\ga
10^{23}$cm$^{-2}$ assuming either an intrinsic power law with exponent
1.9 or 1.7, consistent with the optical classification of very
obscured AGN.

In the total sample with spectroscopic identification 7
sources are classified as stars (6.6\%), 11 as normal
galaxies (10.4\%), 38 are identified as obscured AGN
(35.8\%), and 50 as unobscured AGN (47.2\%). These fractions
are similar to the findings of other X-ray surveys, as shown
in Table~\ref{others}.

The ChaMP Survey \citep{kim04} covers a total of 14
deg$^{2}$. In their first spectroscopy report, 6 $Chandra$
fields were covered to a depth of $r\simeq 21$ mag
\citep{green04}. In order the compare with their results, we
applied our classification scheme to their data. Namely,
narrow-line and absorption-line galaxies with
$L_X>10^{42}$~ergs~s$^{-1}$ were classified as obscured AGN,
sources with broad lines as unobscured AGN, while the
remaining extragalactic sources were classified as galaxies.
The main reason for the discrepancies between their source
mix and ours (Table~\ref{others}) is the optical magnitude
cut for spectroscopy, $\sim 2$ magnitudes brighter than
CYDER, which explains why their sample is clearly dominated
by unobscured AGN, the optically brighter X-ray emitting
sources.

In Table~\ref{others}, our sample is also compared to both
the Chandra Deep Fields North \citep{brandt01} and South
\citep{giacconi02}, each covering $\sim 0.1$ deg$^{-2}$. In
the first case, we use the spectroscopic follow-up of X-ray
sources by \citet{barger03}, which is 87\% complete for
sources with $R<24$ mag. Here, our classification scheme was
applied directly to their data, finding that a low number of
unobscured AGN was found, which can be explained by the
optical nature of the sources selected for spectroscopic
follow-up. Also, a larger number of galaxies relative to
other surveys can be seen. This can be explained by the very
deep X-ray coverage in the CDF-N, which allows for the
detection of a large number of sources with low
$f_{X}/f_{\rm opt}$ and high spatial density, like
non-active galaxies.

In the CDF-S, our results were compared with the
spectroscopic identifications of X-ray sources from
\citet{szokoly04}. In this case, spectra were obtained for
168 X-ray sources and identifications are 60\% complete for
sources with $R<24$ mag. Compared to the CYDER survey, the
source composition is similar, even though a larger number
of X-ray normal galaxies is found in the CDF-S, as expected
given its fainter X-ray sensitivity. However, the fractions
of obscured to unobscured AGN are similar (within $\sim
10$\%) which can be explained by the similarities in the
spectroscopic follow-up programs, since both CYDER and CDF-S
are $\sim 50$\% complete for X-ray sources with $R<24$ mag.

\section{Discussion}
\label{disc}

The CYDER survey is located in an intermediate regime in terms of area
coverage and sensitivity. A critical step in understanding the
properties of the X-ray population is the existence of extensive
follow-up at other wavelengths. In particular, optical spectroscopy
plays a key role, allowing us to determine redshifts and to identify
the origin of the X-ray emission. Therefore, most X-ray surveys are
limited by their ability to obtain spectroscopic identifications for a
large fraction of the sources, hopefully without biasing the
sample. In the case of the CYDER survey, we used 8m class telescopes
in order to extend the spectroscopic coverage to fainter optical
magnitudes, namely to $R\simeq 24$ mag.

From Table~\ref{others}, it is clear that the kind of X-ray
sources identified in surveys depends directly on the depth
of the optical spectroscopy follow-up. For example,
unobscured AGN are bright in the optical bands, therefore in
surveys with shallow optical follow-up mostly unobscured AGN
are detected (e.g.,ChaMP). On the other hand, deep X-ray
coverage, together with an extensive spectroscopy campaign
based mostly on the Keck 10-m telescopes, allows the CDF-N to
detect more faint optical counterparts. Therefore, the
population in very deep surveys is dominated by normal
galaxies to the CDF-N depths and obscured AGN in the CDF-S
range.

In our survey, a total of 50 (47.2\%) broad-line AGN were
detected. While all of them have a hard X-ray luminosity
$L_X>10^{42}$~ergs~s$^{-1}$, two thirds of them have
$L_X>10^{44}$~ergs~s$^{-1}$ and therefore are classified as
quasars. The average redshift for the broad line sample is
$<z>\sim 1.82$, which is much higher than the value found
for the remaining X-ray sources. This is clearly explained
by the greater optical brightness of unobscured AGN relative
to other X-ray emitters.

\begin{figure}
\figurenum{15}
\includegraphics[angle=270,scale=0.7]{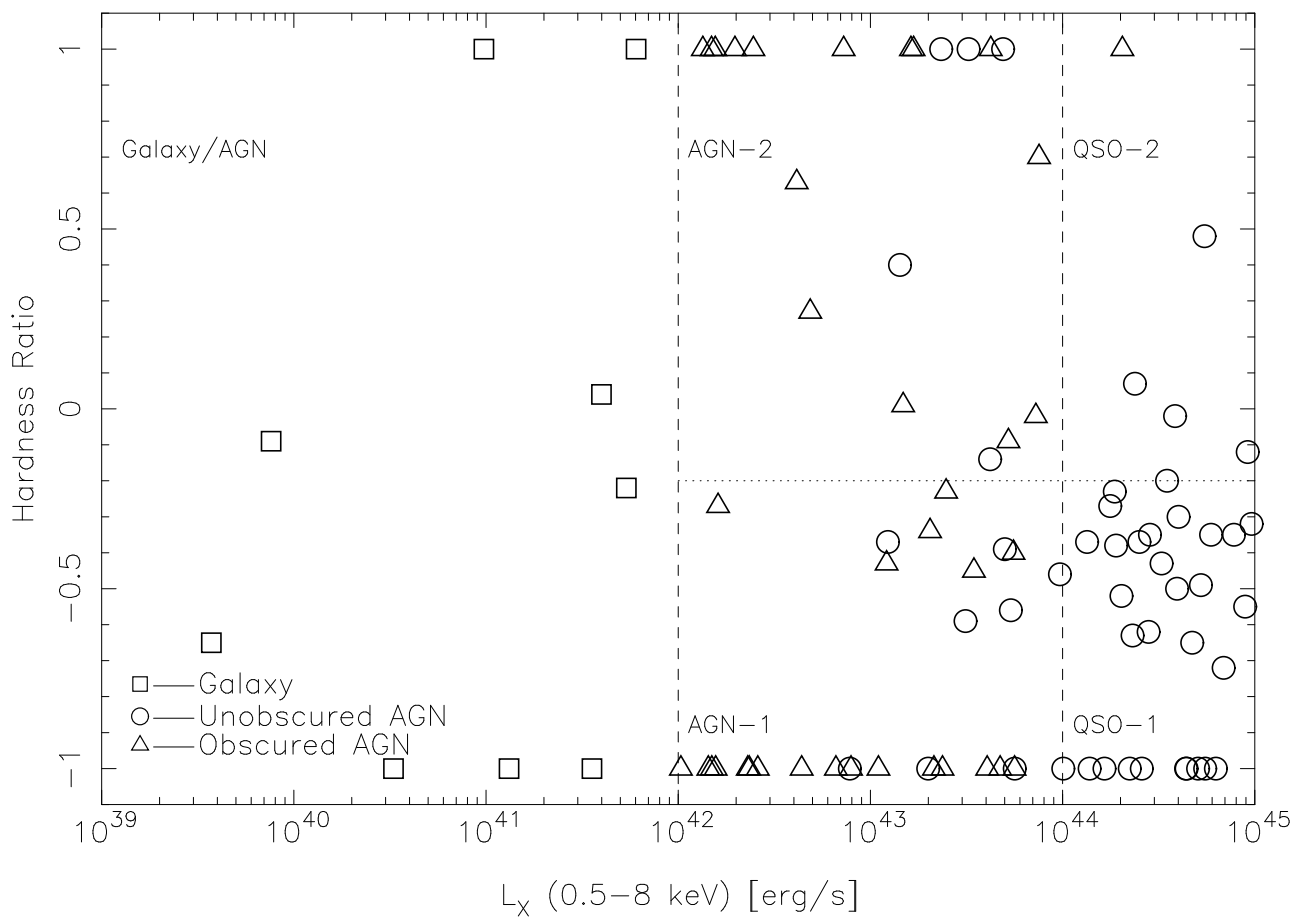}
\caption{HR versus total (0.5-8 keV) X-ray luminosity. Symbols are the same as in Fig~\ref{vi_z}. 
Dashed lines separate galaxies and AGN at
$L_X>10^{42}$~ergs~s$^{-1}$ and ``Quasars'' from lower
luminosity AGN at $L_X=10^{44}$~ergs~s$^{-1}$. The
classification scheme based on the X-ray spectral properties
using HR=-0.2 to separate obscured and unobscured AGN for
sources with $L_X>10^{42}$~ergs~s$^{-1}$ (dotted line) can
be compared to the scheme used in this paper based on the
optical spectrum and X-ray luminosity, showing that in
general obscured AGN have the X-ray hardest spectra.}
\label{hr_lx}
\end{figure}

Using a combination of HR and X-ray luminosity together with
optical spectroscopy is very useful for classifying X-ray
sources \citep{szokoly04}. In Figure~\ref{hr_lx}, the HR
versus hard X-ray luminosity diagram is presented. In this
case we used a HR=-0.2 in AGN-dominated sources rather than
the optical spectra to separate obscured and unobscured AGN,
which is equivalent to an effective column density
$N_H\simeq 4\times 10^{21}$~cm$^{-2}$ for spectral index
$\Gamma=1.9$ or $N_H\simeq 3\times 10^{21}$~cm$^{-2}$ for
$\Gamma=1.7$, so this is a conservative cut to the number of
obscured AGN. Also, quasar-like sources are distinguished
from other X-ray sources using $L_X>10^{44}$~ergs~s$^{-1}$
as a dividing line. Except for one source described in
\S~4.4, all the quasars have broad emission lines in their
optical spectrum. Most sources that show broad emission
lines have HR$<-0.2$, meaning that they have little or no
absorption in X-rays, consistent with their unabsorbed AGN
optical spectrum. For non-AGN dominated X-ray emission, no
correlation is found between HR and X-ray luminosity. Also,
these sources do not have a characteristic HR value and very
hard or soft sources can be found.  This X-ray emission is
expected to be mostly from high-mass X-ray binaries and
type-II supernova remnants in spiral galaxies while for
elliptical galaxies the X-ray emission is most likely
dominated by hot gas with some contribution from low mass
X-ray binaries. Therefore, given the wide range of different
X-ray emitter classes together with the lower luminosity,
which leads to lower fluxes and therefore larger errors in
the HR measurements, can explain why there is no clear
correlation between HR and X-ray luminosity and there is no
characteristic HR value for low-luminosity, non-AGN X-ray
emitters.

\begin{figure} 
\figurenum{16}
\includegraphics[angle=270,scale=0.7]{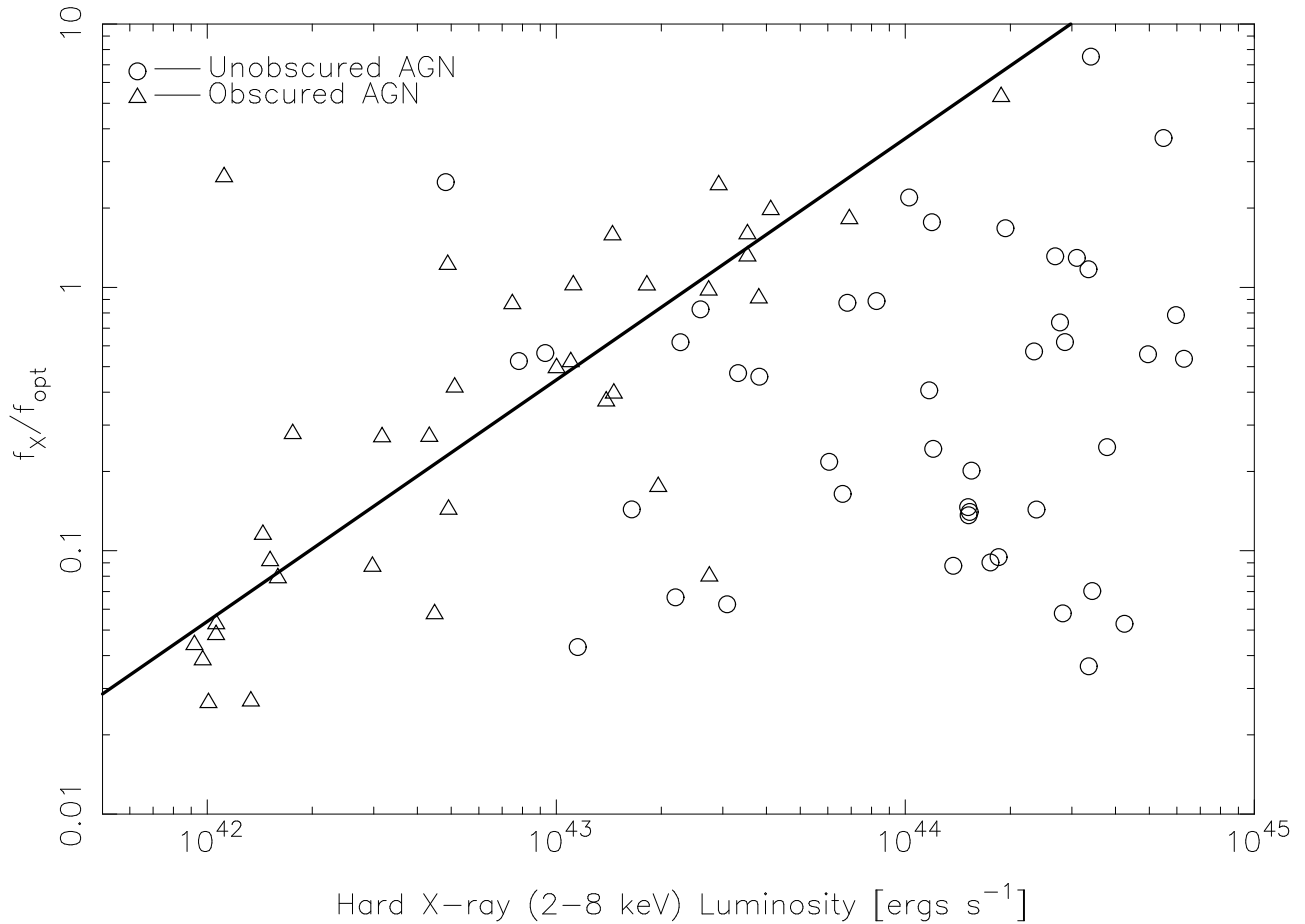} 
\caption{Hard X-ray to optical (measured in the observed frame $V$-band) 
flux ratio versus hard X-ray luminosity for sources with
$L_X>10^{42}$~ergs~s$^{-1}$ (i.e., AGN dominated). While sources
optically classified as broad line AGN ({\it circles}) are scattered
over the $L_X>10^{42}$~ergs~s$^{-1}$ portion of this diagram, for
obscured AGN (narrow emission lines in the spectrum; {\it triangles})
we can observe a rough correlation between $f_X/f_{\rm opt}$ and
$L_X$, namely sources with higher luminosity have larger values of
$f_X/f_{\rm opt}$. The solid line shows the minimum $\chi^2$ fit to
these data. The existence of this correlation can be explained if most
of the optical emission for obscured AGN comes from the host galaxy,
which would be roughly independent of the luminosity of the AGN.}
\label{fxopt_lx} 
\end{figure}

For AGN-dominated sources, the relation between the
$f_X/f_{\rm opt}$ ratio and Hard X-ray luminosity ($L_X$) is
investigated in Figure~\ref{fxopt_lx}. For sources
classified optically as unobscured AGN there is no
correlation between $f_X/f_{\rm opt}$ and X-ray luminosity,
while for obscured AGN there is a clear correlation in the
sense that obscured sources with lower X-ray luminosity have
lower $f_X/f_{\rm opt}$ while the hard X-ray sources with
large X-ray luminosity also have systematically larger
values of $f_X/f_{\rm opt}$. This effect can be explained if
the optical light detected in obscured AGN is dominated by
the emission from the host galaxy (e.g.,
\citealp{treister04b}), that is nearly independent from the
AGN luminosity. Therefore, for obscured sources that are
luminous in X-rays, we can expect a larger $f_X/f_{\rm opt}$
ratio, as observed in our sample. Performing a linear fit to
the observed sample of sources optically classified as
obscured AGN we obtain a correlation at $\sim 2\sigma $
significance using the minimum $\chi^2$ test, with best-fit
parameters given by
\begin{equation}
\log L_X=-39.79(\pm 4.04) +0.917(\pm 0.094)\log (f_X/f_{\rm opt})
\end{equation}
\noindent
This correlation is shown by the solid line in
Figure~\ref{fxopt_lx}. This trend can also be observed at
the same significance level if the $I$-band optical flux is
used instead. This can be explained since the $V$ band is
bluer and therefore it is more affected by dust obscuration,
while in the $I$ band the host galaxy is more luminous, and
therefore in both cases the host galaxy emission dominates
over the AGN optical radiation. A similar relation between
$f_X/f_{\rm opt}$ and X-ray luminosity for obscured AGN was
found by \citet{fiore03} in the High Energy Large Area
Survey (HELLAS2XMM). Even though they used the $R$ band to
calculate the optical luminosity, the correlations are
similar.

\begin{figure}
\figurenum{17}
\includegraphics[angle=270,scale=0.7]{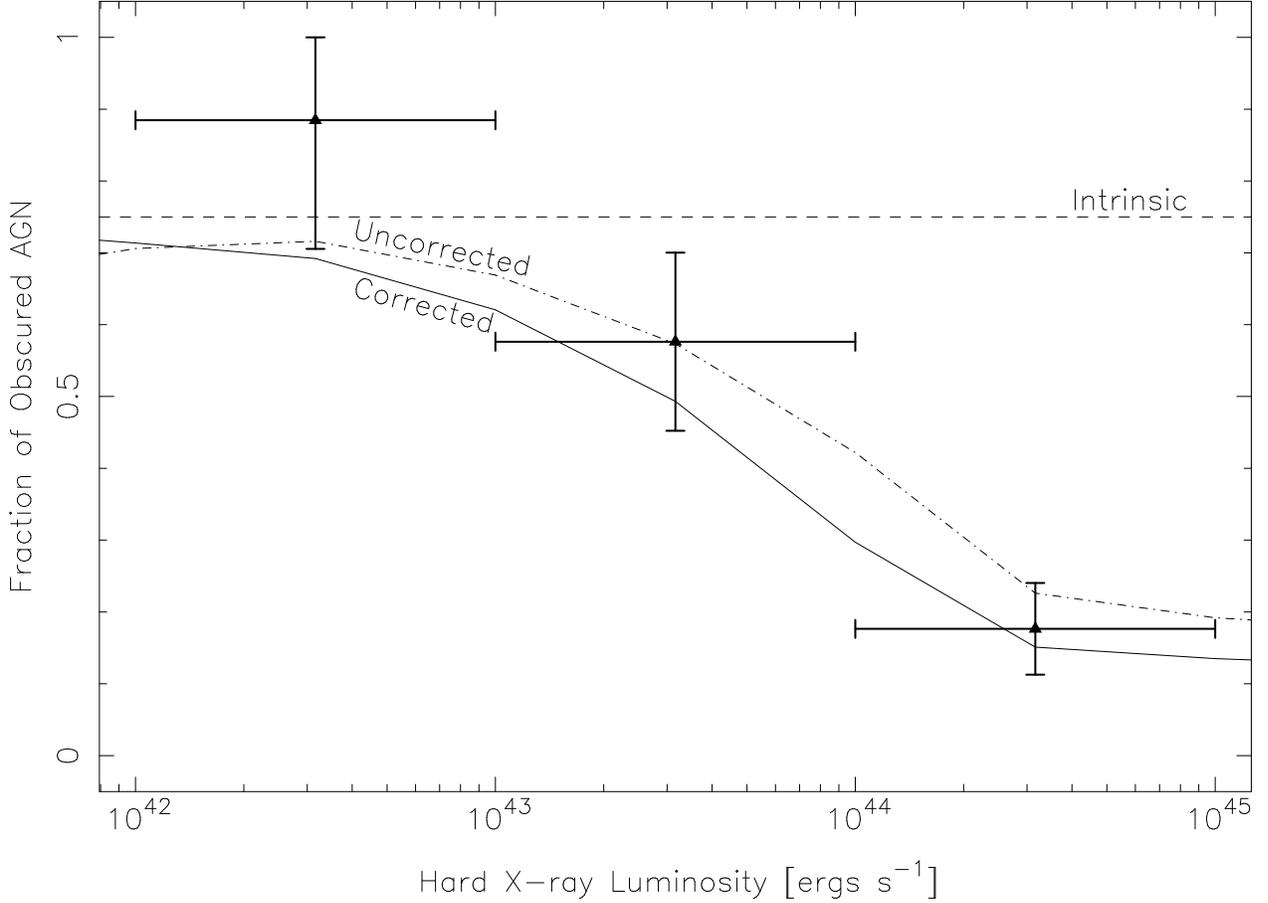}
\caption{Fraction of objects optically classified as obscured AGN versus total 
AGN in $\Delta \log(L_X)=1.0$ bins combining the hard X-ray
sources in the CYDER survey with the sources detected in the
GOODS-S field with spectroscopic identification from
\citet{szokoly04} in order to obtain a larger sample. The
decrease in the number of obscured AGN with X-ray luminosity
can be clearly seen in this figure. ({\it Dot-dashed line})
shows the predicted correlation using the models of
\citet{treister04b} that assumed a constant, fixed, obscured
to total AGN ratio of 3:4 ({\it dashed line}) if only
objects with optical magnitude $R\la 24$ mag (i.e., the
optical cut for spectroscopy) are considered and the effects
of obscuration and k-correction are not taken into account
to calculate the X-ray luminosity. {\it Solid line} shows
the predicted correlation if the intrinsic hard X-ray
luminosity in the model is corrected for obscuration and
redshift effects. From these results, we can see that the
observed correlation can be explained as a selection effect
caused by the need for spectroscopic identification in order
to calculate luminosities.}
\label{frac_lx}
\end{figure}

Given the difficulties in finding obscured AGN at $z>1$, we
are not able to disentangle a dependence of the obscured to
unobscured AGN numbers ratio with redshift from the strong
selection effects on the sample. However, from
Figure~\ref{l_dist} there is some indication that this ratio
can depend on the observed X-ray luminosity. In order to
investigate this effect in more detail, in
Figure~\ref{frac_lx} the fraction of obscured to all AGN is
shown as a function of hard X-ray luminosity combining the
hard X-ray sources detected in the CYDER survey with 77 AGN
with $L_X>10^{42}$~ergs~s$^{-1}$ located in the GOODS-S
field with identifications and redshifts reported by
\citet{szokoly04} in order to increase the number of X-ray
sources in each bin. This figure clearly shows that a
dependence of the fraction of obscured AGN with X-ray
luminosity can be observed. A similar trend was first
observed by \citet{lawrence82} and is consistent with the
relation reported by \citet{ueda03} and \citet{hasinger03}.

In other to further investigate this observed correlation,
and to determine if it can be explained by selection
effects, we used the AGN population models of
\citet{treister04b}. Originally used to predict the AGN number
counts in any wavelength from far infrared to X-rays, the
\citet{treister04b} model is based on the \citet{ueda03}
luminosity function and its luminosity-dependent density
evolution in which the intrinsic $N_H$ distribution comes
from a very simple unified model in which the intrinsic
obscured to unobscured AGN ratio is set to be 3:1. The AGN
spectral energy distribution is modeled based on three
parameters, namely the intrinsic X-ray luminosity of the
central engine, the neutral hydrogen column density in the
line-of-sight and the redshift of the source in order to
compare fluxes in one wavelength to another. Even though
this model was applied to the GOODS survey, it can be
applied to any other X-ray survey if the proper flux limit
and area coverage are used. Given that the luminosity
function and AGN SED library in this model are fixed, there
is no free parameter to adjust.

In Figure~\ref{frac_lx} we show the predicted correlation
between the fraction of obscured to all AGN and hard X-ray
luminosity for sources with $R\la 24$ mag (i.e., the optical
flux limit for spectroscopy) both for intrinsic (dot-dashed
line) and observed (i.e., adding the effects of obscuration
and K-correction; solid line) X-ray luminosity. In both
cases, a decrease in the fraction of obscured to all AGN is
observed with increasing luminosity, even when the intrinsic
ratio is fixed and set to 3:4. Therefore, this observed
correlation can be explained as a selection effect since for
obscured AGN their lower optical flux makes them harder to
detect in spectroscopic surveys, in particular at higher
redshifts where most of the more luminous AGN are located.

A crude way to estimate the intrinsic neutral hydrogen
column density ($N_H$) in the line of sight is based on the
measured HR. In order to estimate $N_H$, we assumed that the
intrinsic X-ray spectrum of an AGN can be described by a
power law with photon index $\Gamma=1.9$ (e.g.,
\citealp{nandra94,nandra97,mainieri02}). We then generated a conversion
table using XSPEC \citep{arnaud96} to calculate the expected
HR for $N_H$ in the range $10^{20}-10^{24}$~cm$^{-2}$ and
redshifts from $z=0$ to $z=5$. The spectral response of the
ACIS camera was considered in this calculation. Also, the
amount of Galactic absorption in each field as calculated
based on the observations by \citet{stark92} was added,
since all the X-ray emission from extragalactic sources
passes through the intergalactic medium of our galaxy. Then,
using this conversion table, the observed HR can be
translated into a $N_H$ value, taking into account the
redshift of the source. Even though for individual sources
this method to estimate $N_H$ may not be very accurate,
given the uncertainties associated to perform spectral
fitting based on only two bins, these individual
uncertainties average out in the distribution.

Given that the ACIS camera is more sensitive in the soft
X-ray band, we decided to exclude sources not detected in
the hard band, in order to use only the sources for which
the HR can give a reasonable idea of the X-ray spectrum. For
AGN-dominated sources (i.e. $L_X>10^{42}$~ergs~s$^{-1}$),
this choice eliminates 34\% of the sources. By cutting the
sample to the sources detected in the hard band, a similar
fraction of objects optically classified as obscured and
unobscured AGN are removed from the sample and therefore we
do not expect a significant bias introduced by this choice
that on the other hand, allows a more precise statistical
analysis, since a definite flux limit can be used. Also,
sources dominated by AGN emission in X-ray have hard
spectrum so if only sources detected in the hard band are
considered, the contamination by non-AGN X-ray emitters is
reduced. Therefore even sources detected with high
significance in the soft band and not detected in the hard
band are removed from the following analysis.

\begin{figure}
\figurenum{18}
\includegraphics[angle=0,scale=0.7]{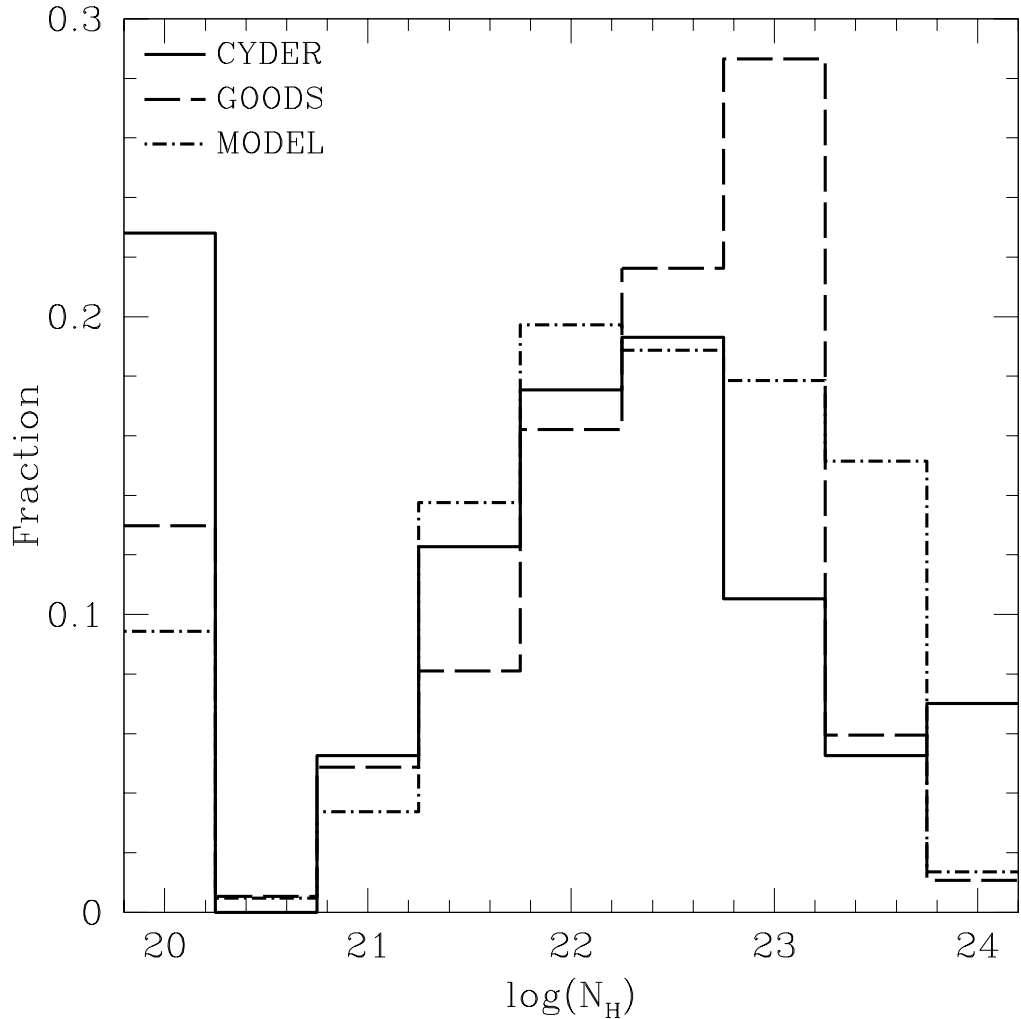}
\caption{Neutral hydrogen column density ($N_H$) distribution 
deduced for X-ray sources with measured spectroscopic
redshift and detected in the hard band. The value of $N_H$
was calculated from the HR assuming an intrinsic power-law
spectrum with exponent $\Gamma=1.9$ ({\it solid line}),
typical for AGN activity, and the spectral response of the
ACIS camera. The redshift of the sources was taken into
account to calculate the intrinsic amount of absorption in
the X-ray spectrum. The $N_H$ distribution for sources in
the GOODS survey ({\it dashed line}), as calculated by
\citet{treister04b} shows that in CYDER absorbed X-ray
sources with $N_H>3\times 10^{22}$~cm$^{-2}$ are
preferentially missed, but they appear in deeper
surveys. {\it Dot-dashed line}: Models from
\citet{treister04b} adapted to the CYDER total area and flux
limits. While the general agreement between can be
considered good, a clear disagreement at the high $N_H$ end
can be observed. This effect was also reported by
\citet{treister04b} based on the GOODS data and can be
explained by the incompleteness in the sample with
spectroscopic redshifts (required to calculate $N_H$) since
highly obscured sources are also the faintest in the optical
bands. The disagreement in the low $N_H$ end is caused by
the presence of sources with a soft-excess in the observed
sample.}
\label{nh_dist}
\end{figure}

The $N_H$ distribution for the sources in the reduced sample
is presented in Figure~\ref{nh_dist}. While a significant
number of sources, 23\%, have $N_H$ values consistent with
no absorption (plotted at $N_H=10^{20}$~cm$^{-2}$), some
sources present moderate to high levels of absorption, with
$N_H>10^{23}$~cm$^{-2}$ ($\sim 12$\%). The $N_H$
distribution for the X-ray sources in the GOODS survey
\citep{dickinson02,giavalisco04}, which overlaps with the
Chandra Deep Fields North and South, was calculated
previously following a similar procedure by
\citet{treister04b}. The results of this calculation are
also presented in Figure~\ref{nh_dist} scaled to the number
of sources in the CYDER survey. Comparing the results from
these two surveys with the predictions for the intrinsic
$N_H$ distribution based on a simple AGN unified model and
the \citet{ueda03} luminosity function made by
\citet{treister04b}, we found that the obscuration bias is
more important for CYDER than for GOODS, meaning that
sources with $N_H>3\times 10^{22}$~cm$^{-2}$ are
preferentially missed in the CYDER survey, since obscuration
makes them fainter even in the hard X-ray bands.

\begin{figure}
\figurenum{19}
\includegraphics[angle=0,scale=0.7]{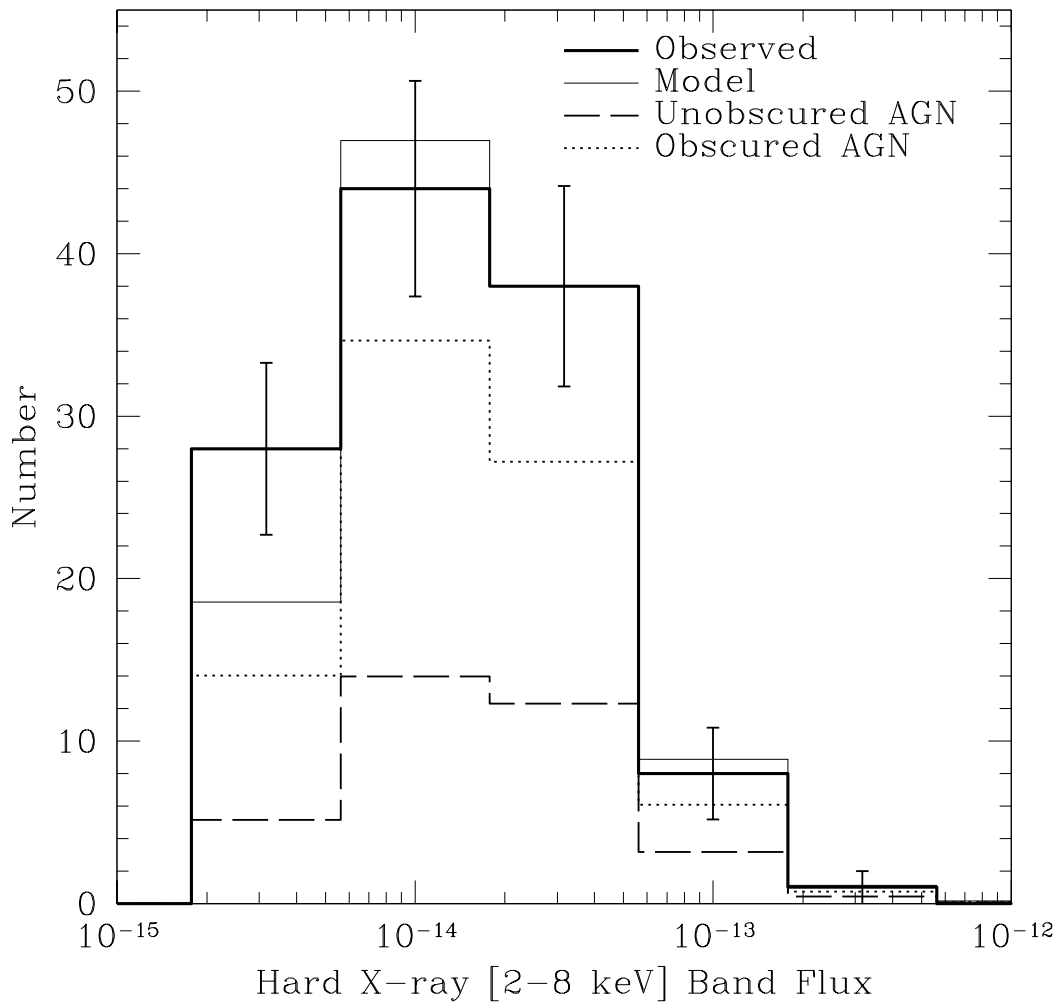}
\caption{Hard X-ray (2-8 keV) flux distribution for sources detected 
in the CYDER fields ({\it heavy solid line}) and predicted
using the simple unified models of \citet{treister04b}; {\it
solid line}. Predicted contributions by unobscured (type 1)
AGN ({\it dashed line}) and obscured (type 2) AGN ({\it
dotted line}) are also shown. The agreement between the
predicted and observed distributions is good, with a K-S
confidence level to accept the null hypothesis of $\sim
96$\%.}
\label{dist_h}
\end{figure}

Using the AGN number counts predictions by
\citet{treister04b} adapted to the CYDER flux limits and
area coverage, the observed hard X-ray flux distribution is
compared to the predictions by this model
(Figure~\ref{dist_h}).  When this sample is compared to the
predictions by the \citet{treister04b} model the results are
very encouraging, showing a very good agreement
characterized by a K-S confidence level to accept the null
hypothesis of $\sim 96$\%. Using this model, in
Figure~\ref{dist_h} the predicted contribution by unobscured
(type 1; {\it dashed line}) and obscured (type 2; {\it
dotted line}) AGN are shown.

While in the \citet{treister04b} model the intrinsic ratio
of obscured to unobscured AGN is 3:1 (using
$N_H=10^{22}$~cm$^{-2}$ as the dividing point), the
prediction for the CYDER X-ray sample is a ratio of 2.35:1
when the survey flux limit in the X-ray bands is
considered. This is consistent with the claim that sources
with $N_H>3\times 10^{22}$~cm$^{-2}$ are preferentially
missed in the CYDER X-ray sample. However, this ratio should
be compared to the value of 0.76:1 obtained previously using
optical spectroscopy to separate obscured and unobscured
AGN. This significant reduction in the relative number of
sources classified as obscured AGN can be explained by the
optical magnitude cut introduced when optical spectroscopy
is used. In the case of the CYDER multiwavelength follow-up,
only sources with $V<25$ mag have optical spectroscopy, and
the completeness level decreases strongly with decreasing
optical flux (Figure~\ref{v_dist}). Since obscured AGN are
in general faint optical sources
(e.g. \citealp{alexander01,koekemoer02,treister04b}), they
are harder to identify using spectroscopy, which cause their
relative number to decrease when compared to other X-ray
sources that are brighter in the optical bands, like
unobscured AGN.

\section{Conclusions}
\label{conc}

We presented here the first results from the multiwavelength
study of the X-ray sources in the CYDER survey. In this
work, we studied the optical and X-ray properties of 267
sources detected in 5 fields observed by $Chandra$ and
available in the archive, covering a total of $\sim
0.1$~deg$^{-2}$ and spanning a flux range of
$10^{-15}-10^{-13}$~ergs~cm$^{-2}$~s$^{-1}$.

The X-ray flux distribution of CYDER sources follows a $\log
N-\log S$ relation, both cumulative and differential, that
is consistent with the observations in existing X-ray
survey. The cumulative $\log N-\log S$ distribution is
consistent with the observations of \citet{ueda03}, while
the differential $\log N-\log S$ is in good agreement with
the distribution derived by the SEXSI survey
\citep{harrison03}. This implies that there are not
significant variations in this sample compared to other
existing surveys, and therefore that the results can be
directly compared.

In general, sources optically classified as obscured AGN
have redder optical colors than unobscured AGN and are
closer to the colors of normal galaxies, as expected from
the unification model of AGN. Also, a correlation between
$f_X/f_{\rm opt}$ and hard X-ray luminosity is observed in
the sample of sources optically classified as obscured AGN.

The ratio of obscured AGN seems to be changing as a function
of X-ray luminosity, in the sense that for more luminous
sources the ratio of obscured to unobscured AGN is lower
than for less luminous objects. However, this relation can
be explained as a selection effect since obscured AGN are
fainter in the optical bands and therefore harder to
identify for spectroscopic surveys. In fact, the observed
correlation can be reproduced using the \citet{treister04b}
models that have a fixed intrinsic ratio of 3:4 if an
optical cut of $R\la 24$ mag (i.e., the magnitude limit for
spectroscopy) is used.

The $N_H$ distribution for sources in the CYDER survey is
consistent with the predicted distribution by
\citet{treister04b} assuming a torus geometry for the
obscuring material once selection effects are
accounted. This implies that X-ray surveys are subject to
significant incompleteness for sources with large amounts of
absorption. In the particular case of the CYDER survey this
incompleteness is important for sources with $N_H>3\times
10^{22}$~cm$^{-2}$. However, once these selection effects
are accounted for, the observed hard X-ray flux distribution
is consistent with the predictions of the models of
\citet{treister04b}.

\acknowledgements

ET would like to thank the support of Fundaci\'on Andes,
Centro de Astrofisica FONDAP and the Sigma-Xi foundation
through a Grant in-aid of Research. This material is based
upon work supported by the National Science Foundation under
Grant No. AST-0201667, an Astronomy \& Astrophysics
Postdoctoral Fellowship awarded to E. Gawiser. We thank the
anonymous referee for a very careful review and a
constructive report that improved the presentation of this
paper. We would like to thank the help of Steve Zepf, Rafael
Guzman and Maria Teresa Ruiz in the original design of this
survey. We also thank the assistance during the observations
provided by the staff at Las Campanas Observatory, Cerro
Tololo International Observatory and Cerro Paranal.


\newpage

\begin{deluxetable}{cclcccccccc}
\tablecolumns{22}
\tabletypesize{\footnotesize}
\rotate
\tablewidth{0pc}
\tablecaption{CYDER Fields}
\tablehead{
\colhead{ID} & \colhead{Obs. ID} &\colhead{Target} & \colhead{Exposure} & \colhead{Mode} &
\colhead{RA} & \colhead{Dec} & \colhead{Gal} & \colhead{Gal N$_H$} & \colhead{Eff. Area} &\colhead{Hard Flux Lim}\\
\colhead{} & \colhead{} &\colhead{} & \colhead{Time (ks)} & \colhead{} & \colhead{(J2000)} & \colhead{(J2000)} &
 \colhead{Lat.} & \colhead{($10^{20}$~cm$^{-2}$)} & \colhead{(arcmin$^2$)} &\colhead{(erg~cm$^{-2}$s$^{-1}$)}}
\startdata
C2 & 921 & HCG 62 & 49.15 & ACIS-S & 12\fh 53\fm 05\fs7 & -09\fdg 12\farcm 20\farcs0 & 53.66$^\circ$ & $3.11$ & 124.0 & $7.9\times 10^{-15}$\\ 
C5 & 866 &Q1127-145 & 30.16 & ACIS-S & 11\fh 30\fm 07\fs1 & -14\fdg 49\farcm 27\farcs0 & 43.64$^\circ$ & $4.04$ & 0.0 & $1.6\times 10^{-14}$ \\
D1 & 905 &HCG 90 & 50.16 & ACIS-I & 22\fh 02\fm 04\fs0 & -31\fdg 58\farcm 30\farcs0 & -53.08$^\circ$ & $1.64$ & 114.24 & $1.44\times 10^{-15}$\\
D2 & 861 &Q2345+007 & 75.15 & ACIS-S & 23\fh 48\fm 19\fs6 & 00\fdg 57\farcm 21\farcs0 & -58.07$^\circ$ & $3.77$ & 62.0 & $1.33\times 10^{-15}$ \\
D3 & 796 &SBS 0335-052 & 60.51 & ACIS-I & 03\fh 37\fm 44\fs0 & -05\fdg 02\farcm 39\farcs0 & -44.69$^\circ$ & $4.91$ & 120.0 & $3.59\times 10^{-15}$
\enddata
\label{fields}
\end{deluxetable}

\begin{deluxetable}{ccccccc}
\tablecolumns{14}
\tablewidth{0pc}
\tablecaption{Summary of Optical Observations}
\tablehead{
\colhead{Field} & \colhead{Instrument} & \colhead{Date} & \colhead{Filters} & \colhead{Exposure} & \colhead{Seeing} & \colhead{Lim.}\\
\colhead{}      & \colhead{}           & \colhead{}     & \colhead{}        & \colhead{Time}  & \colhead{} & \colhead{Magnitude (5$\sigma$)}
}
\startdata
C2 & CTIO/MOSAIC & 03/11/2000 & $V$ & 6000s & 1.1'' &  26.5\\
C2 & CTIO/MOSAIC & 03/11/2000 & $I$ & 1500s & 1.0'' & 24.9 \\
C5 & CTIO/MOSAIC & 03/12/2000 & $V$ & 6000s & 1.1'' & 26.5\\
C5 & CTIO/MOSAIC & 03/12/2000 & $I$ & 1500s & 1.0'' & 24.9\\
D1 & CTIO/MOSAIC & 08/22/2001 & $V$ & 6600s & 1.0'' & 26.7\\
D1 & CTIO/MOSAIC & 08/22/2001 & $I$ & 1800s & 0.9'' & 25.1\\
D2 & CTIO/MOSAIC & 08/23/2001 & $V$ & 1500s & 1.4'' & 25.5\\
D2 & CTIO/MOSAIC & 08/23/2001 & $I$ & 5100s & 2.0'' & 24.8\\
D3 & CTIO/MOSAIC & 10/10/2002 & $V$ & 3000s & 1.2'' & 26.1\\
D3 & CTIO/MOSAIC & 10/10/2002 & $I$ & 1200s & 1.1'' & 24.6\\
\enddata
\label{opt_sum}
\end{deluxetable}

\begin{deluxetable}{cccccc}
\tablecolumns{12}
\tablewidth{0pc}
\tablecaption{Summary of Spectroscopic Observations}
\tablehead{
\colhead{Field} & \colhead{Instrument} & \colhead{Date} & \colhead{Masks} & \colhead{Exp. Time} & \colhead{Seeing} \\
\colhead{}      & \colhead{}           & \colhead{}     & \colhead{}        & \colhead{per Mask}  & \colhead{}
}
\startdata
C2 & VLT/FORS2 & 02/14/2002 & 3 & 6300s & 1.0'' \\ 
C5 & VLT/FORS2 & 02/15/2002 & 1 & 8000s & 1.2'' \\ 
D1 & Baade/LDSS-2 & 10/03-04/2002 & 3 & 6300s & 0.7'' \\ 
D1 & Baade/LDSS-2 & 10/03-04/2002 & 3 & 6900s & 1.0'' \\ 
D2 & Baade/LDSS-2 & 10/03-04/2002 & 2 & 7200s & 0.9'' \\ 
D3 & VLT/FORS2 & 10/29-31/2003 & 3 & 9200s & 0.5'' \\ 
D3 & VLT/FORS2 & 10/29-31/2003 & 6 & 9200s & 0.6'' \\ \enddata
\label{spec_sum} 
\end{deluxetable}

\begin{deluxetable}{lcccccccccccccccc}
\tablecolumns{17}
\tabletypesize{\scriptsize}
\rotate
\tablewidth{0pc}
\tablecaption{Catalog of X-ray sources in the CYDER fields.}
\tablehead{
\colhead{Name} & \colhead{RA} & \colhead{Dec} & \colhead{Offset} &\colhead{SB} & \colhead{HB} & \colhead{SB Error} & \colhead{HB Error} 
& \colhead{HR} & \colhead{V Mag.} & \colhead{I Mag.} & \colhead{SB flux} & \colhead{HB flux} & \colhead{z} & \colhead{SB Lum.} & \colhead{HB Lum.} & 
\colhead{Source}\\
\colhead{CXOCY-} & \colhead{} & \colhead{}  &\colhead{} &\colhead{Counts} & \colhead{Counts} & \colhead{Counts} & \colhead{Counts} &\colhead{} & \colhead{} & \colhead{}  
& \colhead{erg~cm$^{-2}$s$^{-1}$} & \colhead{erg~cm$^{-2}$s$^{-1}$}
& \colhead{} & \colhead{erg~s$^{-1}$} & \colhead{erg~s$^{-1}$} & \colhead{Class}
}
\startdata
J125302.4-091312&12 53 02.40&-09 13 11.8&1.104&380.3&184.9&20.1&13.8&-0.35& 20.50& 20.46&3.80e-14&7.20e-14& 1.13& 2.62e+44& 4.96e+44& Q \\
J125314.8-091301&12 53 14.82&-09 13 01.2&1.047&  9.1& 52.5& 3.2& 7.3& 0.70& 22.70& 20.70&1.90e-15&3.10e-14& 0.72& 4.24e+42& 6.91e+43& 2 \\
J125314.6-091050&12 53 14.64&-09 10 49.8&-----& 24.5&  8.4& 5.0& 3.0&-0.49&>26.50&>24.90&2.60e-15&4.80e-15& ----& --------& --------& X \\
J125311.1-091118&12 53 11.12&-09 11 17.7&1.027&  0.0& 19.8& 0.0& 4.6& 1.00& 21.34& 19.96&7.00e-16&1.60e-15& 0.48& 5.84e+41& 1.33e+42& 2 \\
J125310.4-091024&12 53 10.41&-09 10 23.6&0.831& 36.5& 20.9& 6.1& 4.7&-0.27& 22.62& 21.19&4.20e-16&8.80e-16& 0.56& 5.06e+41& 1.06e+42& 2 \\
J125306.1-091344&12 53 06.08&-09 13 43.9&-----& 28.6& 32.8& 6.4& 6.1& 0.07&>26.50&>24.90&2.50e-16&1.10e-15& ----& --------& --------& X \\
J125306.0-091316&12 53 06.00&-09 13 16.4&0.912& 48.8& 40.7& 7.7& 6.8&-0.09& 22.60& 20.06&5.70e-15&1.70e-14& 0.72& 1.28e+43& 3.80e+43& 2 \\
J125305.3-090824&12 53 05.34&-09 08 23.6&0.592&136.2& 51.6&11.7& 7.3&-0.45& 20.58& 19.70&1.50e-14&2.10e-14& 0.50& 1.40e+43& 1.96e+43& 2 \\
J125305.0-091339&12 53 05.01&-09 13 39.0&-----& 66.4& 34.7& 9.3& 6.3&-0.31&>26.50&>24.90&7.30e-15&1.30e-14& ----& --------& --------& X \\
J125303.8-090810&12 53 03.83&-09 08 09.7&-----& 60.5& 48.5& 7.9& 7.1&-0.11&>26.50&>24.90&8.10e-15&2.00e-14& ----& --------& --------& X \\
J125303.0-091242&12 53 03.01&-09 12 41.9&0.855& 95.8& 30.3&10.7& 5.9&-0.52& 23.48& 21.80&8.50e-15&1.50e-14& ----& --------& --------& X \\
J125302.9-091058&12 53 02.93&-09 10 57.7&0.651&  0.0& 13.8& 0.0& 3.9& 1.00& 21.77& 20.56&3.60e-16&8.40e-16& 0.38& 1.76e+41& 4.11e+41& G \\
J125301.9-091134&12 53 01.90&-09 11 33.9&0.853& 27.6& 13.6& 5.6& 3.9&-0.34& 22.70& 21.40&2.70e-15&6.30e-15& 0.71& 5.96e+42& 1.39e+43& 2 \\
J125300.9-090941&12 53 00.89&-09 09 41.3&0.904&  0.0& 17.7& 0.0& 4.4& 1.00& 24.85& 22.55&1.10e-15&2.40e-15& 0.97& 5.13e+42& 1.12e+43& 2 \\
J125317.5-091223&12 53 17.55&-09 12 23.3&0.795& 23.6&  9.8& 5.0& 3.3&-0.42& 26.05& 24.56&2.90e-15&4.00e-15& ----& --------& --------& X \\
J125315.2-091424&12 53 15.21&-09 14 24.4&1.280&  0.0& 52.5& 0.0& 7.6& 1.00& 24.05& 22.88&1.50e-15&2.60e-14& 1.15& 1.09e+43& 1.88e+44& 2 \\
J125311.8-091339&12 53 11.84&-09 13 38.6&-----& 35.1& 17.9& 6.3& 4.7&-0.32&>26.50&>24.90&3.30e-15&8.70e-15& 3.05& 2.57e+44& 6.78e+44& Q \\
\enddata
\tablecomments{This table is published in its entirety in the electronic edition of the Astrophysical Journal. A portion is shown here for guidance regarding its form and content.}
\label{catalog}
\end{deluxetable}

\begin{deluxetable}{cccccc}
\tablecolumns{12}
\tablewidth{0pc}
\tablecaption{Source Compositions in X-ray Surveys}
\tablehead{
\colhead{} & \multicolumn{4}{c}{Type} & \colhead{}\\
\cline{2-5}
\colhead{Survey} & \colhead{Stars} & \colhead{Galaxies} & \colhead{Obscured AGN} & \colhead{Unobscured AGN} & \colhead{Ref}
}
\startdata
CYDER & 6.6\% & 10.4\% & 35.8\% & 47.2\% & 1\\
ChaMP & 9.9\% & 14.1\% & 23.9\% & 52.1\% & 2\\
CDF-N & 4.9\% & 43.3\% & 39.4\% & 12.4\% & 3\\
CDF-S & 4.5\% & 17.8\% & 42.7\% & 35.0\% & 4\\
\enddata
\tablerefs{
(1) This paper; (2) \citealp{green04}; (3) \citealp{barger03}; (4) \citealp{szokoly04}
}
\label{others}  
\end{deluxetable}

\newpage

\clearpage

\end{document}